\documentclass[aps,prl,reprint,superscriptaddress]{revtex4-1}

\usepackage{amsmath,amssymb,amsfonts,mathrsfs,bm,dsfont}
\usepackage{slashed}
\usepackage{enumerate}
\usepackage{enumitem} % Customize itemize, see https://ctan.org/tex-archive/macros/latex/contrib/enumitem/
\usepackage[all]{xy}
\usepackage{cleveref} % multiple equation ref, see https://tex.stackexchange.com/questions/314217/how-i-can-refer-multiple-equation-in-latex?rq=1
\usepackage{graphics,color}
\usepackage[dvipsnames]{xcolor}
\usepackage[normalem]{ulem} % for strikethrough lines; usage: \sout{}
\usepackage{tikz}

\allowdisplaybreaks[4] % allow align environment to break the page

\newcommand*\dd{\mathop{}\!\mathrm{d}}
\begin{document}

\author{Xiaodong Hu}
\affiliation{Department of Physics, Boston College, Chestnut Hill, Massachusetts 02467, USA}

\author{Jung Hoon Han}
\affiliation{Department of Physics, Sungkyunkwan University, Suwon 16419, Korea}

\author{Ying Ran}
\affiliation{Department of Physics, Boston College, Chestnut Hill, Massachusetts 02467, USA}

\date{\today}
\title{Supercurrent-induced Anomalous Thermal Hall Effect as a New Probe to\\ Superconducting Gap Anisotropy}

\begin{abstract}
Two-dimensional superconductors have been realized in various atomically thin films such as the twisted bilayer graphene, some of which are anticipated to involve unconventional pairing mechanism. Due to their low dimensionality, experimental probes of the exact nature of superconductivity in these systems have been limited. We propose, by applying a \emph{vertical} supercurrent to a bilayer superconductor where the mirror symmetry is naturally broken by the twisting, there will be anomalous thermal Hall effect induced by the supercurrent that can serve as a sharp probe for the \emph{in-plane} anisotropy of the superconducting gap function. This effect occurs in the \emph{absence} of an external magnetic field and spontaneous breaking of the time-reversal symmetry in the ground state. We derive explicit formulas for the induced thermal Hall conductivity and show them to be significant in the examples of twisted cuprates and twisted FeSe where monolayer superconductivity have already been observed. Though technical challenges still exist, we propose this to be a generic probe of the gap anisotropy in a twisted bilayer superconductor.
\end{abstract}

\maketitle

\textit{Introduction}.---
The discovery of the correlated insulating phases and superconductivity in twisted bilayer graphene (TBG) \cite{cao2018unconventional,cao2018correlated} has spurred intense interest in twisted two-dimensional (2D) heterostructures, leading to the notion of \emph{twistronics} as a new form of electronic device. To understand the (possibly unconventional) superconductivity realized in these systems, intense experimental investigations have been on-going \cite{oh2021evidence,kim2022evidence,lake2022pairing}. While the majority of transport studies on twisted bilayers focuses on electrical transport at the moment, thermal transport has long been established as a powerful and complementary tool for investigating the nature of elementary excitations, particularly in superconductors where ordinary electric transport measurement is ineffective \cite{krishana1997plateaus,chiao2000low,sutherland2003thermal,durst2003weak,zhang2001giant,cvetkovic2015berry}. More recently, quantized thermal Hall conductivity at low temperature became a signature of the topologically ordered ground states in correlated materials \cite{banerjee2018observation,yokoi2021half}. In this work, we show that the thermal Hall response can be a sensitive probe of the {\it gap anisotropy} in twisted bilayer superconductors, or TBS for short. 

The gap function $\varDelta_{\bm{k}}$ of a superconductor has immense implications for the underlying pairing mechanism and the quasiparticle transport. It may be deduced in the angle-resolved photoemission spectroscopy (ARPES) \cite{shen1993anomalously,ding1996angle,zhang2016superconducting}, or through the quasiparticle interference (QPI) imaging in scanning tunneling spectroscopy \cite{hanaguri2007quasiparticle,sprau2017discovery}. Resolving the gap anisotropy in ARPES becomes challenging though for low temperature superconductors where $\varDelta_{\bm{k}}$ is smaller than the experimental resolution. For twisted heterostructures, resolving the momentum space structure within a small moir\'e Brillouin zone (BZ) necessitates the QPI imaging over a formidably large area in the real space. As the demand to resolve the gap structure in TBS grows, limitations of existing experimental probes seem to loom larger. A natural question to ask, at this stage, is whether it is possible to invent a new probe of the gap anisotropy for very small $\varDelta_{\bm{k}}$. Here we propose a supercurrent-induced anomalous thermal Hall effect (SATHE) as one possible way to directly probe the gap anisotropy in TBS. 

\begin{figure}[!ht]
	\centering
	\includegraphics[scale=0.5]{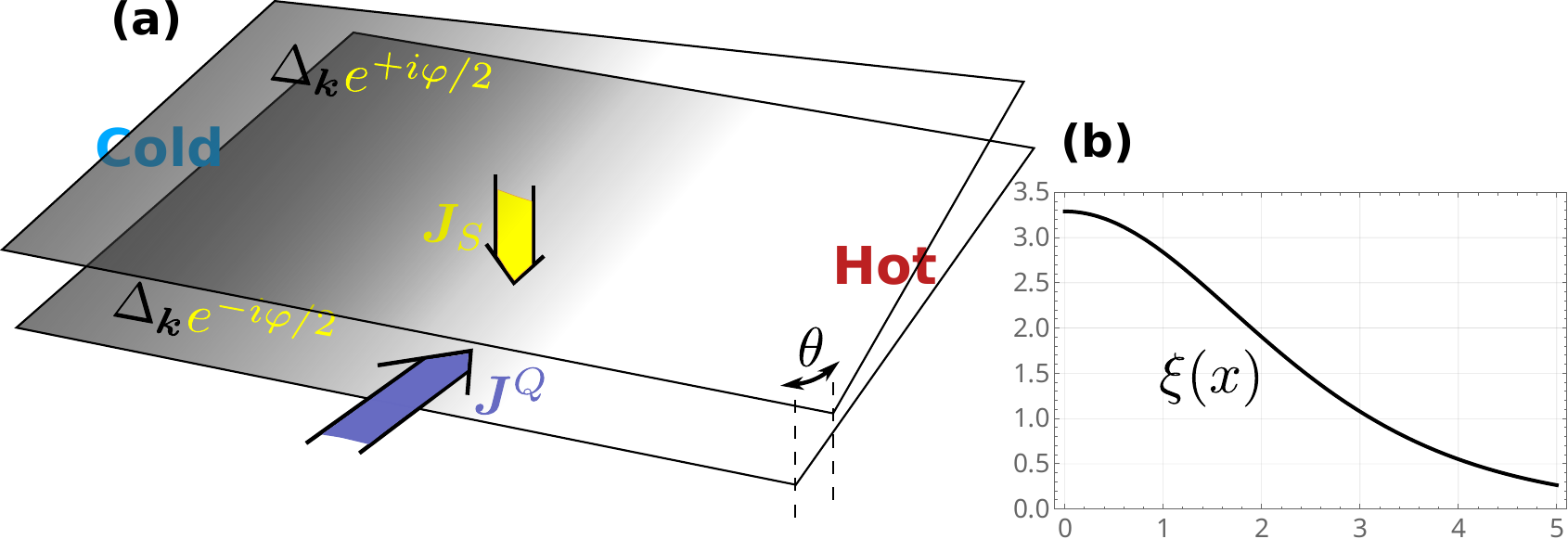}
	\caption{(a) SATHE for bilayer-SC. The vertical supercurrent $\bm{J}_S$, or equivalently, a pairing phase twist $\varphi$ between the two layers of, either a vertical JJ or an intrinsic bilayer superconductor, would induce an in-plane thermal Hall effect in a bilayer superconductor. In the former case, an insulating buffer layer may be present (not shown).  (b) an illustration for the dimensionless function $\xi(x)$ in Eq.(\ref{eq:xi}).}
	\label{fig: setup}
\end{figure}

Fig \ref{fig: setup}.(a) shows the schematic setup for SATHE. The TBS may be a vertical Josephson junction (JJ) formed by stacking two atomically thin superconducting films with a certain twist angle, or an intrinsic twisted bilayer superconductor as in TBG. SATHE is a nonlinear response of heat, created by simultaneously applying a \emph{vertical} supercurrent $\bm J_S$ and an {\it in-plane} temperature gradient, with the resulting {\it transverse in-plane} flow of heat. In contrast to the conventional thermal Hall effect (THE) which occurs when the ground state breaks the time-reversal symmetry (TRS), SATHE can occur for ground states that preserve the TRS. No external magnetic field is required to observe SATHE. 

Our proposal thus differs for most proposals of THE including, for instance, Ref. \cite{can2021high} where the JJ spontaneously breaks TRS in the absence of applied supercurrent. It bear resemblance to the Sodemann-Fu proposal for nonlinear electrical Hall effect \cite{sodemann2015quantum,ma2019observation}, in that in both proposals the \emph{unperturbed} ground state preserves TRS. One can think of the nonlinear electrical Hall effect as arising from the first electric field driving an imbalance of the fermion distribution, and the second electric field is used to probe the Hall response. In our proposal, the supercurrent is employed to drive Berry curvature out of its equilibrium form, then a temperature gradient is applied to induce the thermal Hall response. In this sense, SATHE can be taken as a thermal analogue of the nonlinear Hall effect.

When the supercurrent is not too large, one can linearize the current-phase relation $\bm J_S \propto\varphi$ and SAHTE becomes the perturbative change of the thermal Hall conductivity $\delta \kappa_{xy}$ proportional to the phase twist, $\delta\kappa_{xy}\simeq\varphi\cdot\chi_{\varphi xy}$, with $\chi_{\varphi xy}$ capturing the nonlinear response of the system. Similar to nonlinear Hall effect, the breaking of inversion symmetry is necessary to elicit the desired responses. The observation of SATHE additionally requires the breaking of in-plane and out-of-plane mirror symmetries, which are not conditions normally present in the family of nonlinear Hall effects, but are naturally satisfied in the TBS. The conditions for observing SATHE are therefore not any more stringent than those of other proposed nonlinear Hall effect, at least from the perspective of symmetry requirement.

\textit{Main results}.--- For TBS with a multiband electronic structure, the pairing Hamiltonian is generally $\sum_{a,b} \varDelta_{\bm{k}}^{a,b} c^\dagger_{a,\bm{k}} c^\dagger_{b,-\bm{k}}$, with $a,b$ ranging over the bands. We firstly study the simple situation where only the \emph{intraband pairing} is present: $\varDelta_{\bm{k}}^{a,b}=0$ for $a\neq b$. Assuming that $|\varDelta_{\bm{k}}|$ is much smaller than the energy difference between bands in the normal state,  we find that a small pairing phase difference $\varphi$ between the layers in the TBS induces a change of the Berry curvature and leads to the SATHE formula
\begin{align}
    & \frac{\delta\kappa_{xy}^{\text{intra.}}}{T} = \varphi\cdot \frac{\chi_{\varphi xy}^{\text{intra.}}}{T} =\varphi \cdot\frac{k_B^2}{16\pi^2\hbar} \sum_{\text{FS}}\mathop{\mathrm{sign}}(v_F)\notag\\
	% \frac{\delta\kappa^{\text{intra.}}_{xy}}{\delta\varphi\cdot T}=&\frac{k_B^2}{16\pi^2\hbar} \sum_{\text{FS}}\mathop{\mathrm{sign}}(v_F)\notag\\
	&\qquad\times\oint_{\text{FS}}\dd k_\parallel\,\xi\bigg(\frac{\varDelta_{\bm{k}}(T)}{k_BT}\bigg)\cdot\partial_{k_\parallel} [\langle u_{\bm{k}}|\hat L|u_{\bm{k}}\rangle] .  \label{eq:main_result}
\end{align}
Here $\chi_{\varphi xy}^{\text{intra.}}$ is the intraband transport coefficient for SATHE, $\xi(x)$ is the dimensionless function (see Fig.\ref{fig: setup} (b))
\begin{align}\label{eq:xi}
	\xi(x)\equiv\int_{|x|}^\infty\mathrm{d}x' \sqrt{x'^2-x^2}\frac{x'}{1+\cosh(x')},
\end{align}
and $|u_{\bm{k}}\rangle$ is the Bloch state at the Fermi level in the \emph{normal} state. The dimensionless Hermitian operator $\hat L$ is the generator of Doppler shift due to $\varphi$, and equals $\hat L=\mathop{\mathrm{diag}}\{\mathbf{1},\mathbf{-1}\}$ in the layer space  for a vertical JJ but gets more complicated for an intrinsic bilayer superconductor (see below and Supplemental Material (SM) \cite{SM}). The loop integral is performed over each Fermi surface (FS) and $k_\parallel$ is the counterclockwise tangential momentum at the Fermi surface. The $\mathop{\mathrm{sign}}(v_F)=\pm 1$ characterizes whether the FS is electron-like or hole-like (a single band may host multiple FS's). Clearly if $\varDelta_{\bm{k}}$ is $\bm{k}$-independent, the loop integral over the FS reduces to a total derivative and vanishes identically.

\textit{The $\hat L$ operator and Doppler shift}.---
The normal state tight-binding Hamiltonian for the bilayer can be written in a block form:
\begin{align}
	H_0(\bm{k})=\begin{pmatrix} H_0^t(\bm{k}) & T_\perp(\bm{k}) \\ T^\dagger_{\perp}(\bm{k}) & H_0^b(\bm{k})\end{pmatrix},\label{eq:H_0}
\end{align}
where $t/b$ labels the top/bottom layer, and $T_\perp$ represents the interlayer hopping. In the superconducting state, pairing terms are introduced. Introducing the pairing phase difference $\varphi$ due to the applied supercurrent is equivalent to performing a gauge transformation $H_0\rightarrow U (\varphi) H_0U^\dagger (\varphi)$ (Doppler shift) while keeping the pairing terms unchanged, and $\hat L$ is its generator: $U(\varphi)\equiv e^{-i\frac{\varphi}{4} \hat L}$. For a vertical JJ, $\hat L=\mathop{\mathrm{diag}}\{\mathbf{1},\mathbf{-1}\}$. For an intrinsic superconductor, $\hat L$ depends on the atomic orbital positions normal to the bilayer plane (see SM \cite{SM}).

For a JJ, since $|u_{\bm{k}}\rangle$ is the eigenstate of $H_0(\bm{k})$, $\langle u_{\bm{k}}|\hat L|u_{\bm{k}}\rangle \in [1,-1]$ serve as the indicator of the layer-component of the state. If $T_\perp$ is absent, top/bottom layers decouple and $\langle u_{\bm{k}}|\hat L|u_{\bm{k}}\rangle =\pm 1$, leading to vanishing SATHE according to Eq.(\ref{eq:main_result}), consistent with physical intuitions.

\textit{Derivation of Eq. (\ref{eq:main_result})}.---
Here we show how to derive the SATHE formula of Eq. (\ref{eq:main_result}) while leaving the more technical parts to \cite{SM}. The thermal Hall conductance in a 2D superconductor is  related to the superconducting Berry curvature \cite{qin2011energy,sumiyoshi2013quantum} as
\begin{align}
	\frac{\kappa_{xy}}{T}=-\frac{k_B^2}{2\hbar}\int_{-\infty}^\infty dE \frac{E^2}{(k_B T)^2} \boldsymbol{\sigma}(E) f'(E),\label{eq:kappa}
\end{align}
where $f(E)$ is the Fermi-Dirac function, and 
\begin{align}
	\boldsymbol{\sigma}(E)\equiv -\sum_\mathbf{a} \int_{E_{\bm{k}}<E} \frac{d^2k}{(2\pi)^2} \boldsymbol{\Omega}^\mathbf{a}_{\bm{k}}. \label{eq:sigma}
\end{align}
(Bold fonts are used for quantities related to the superconducting BdG state, to be distinguished from the normal state ones.) Here $\boldsymbol{\Omega}^\mathbf{a}_{\bm{k}}=-2\mathop{\mathrm{Im}}\langle\partial_{k_x} \mathbf{u}^\mathbf{a}_{\bm{k}}|\partial_{k_y} \mathbf{u}^\mathbf{a}_{\bm{k}}\rangle$ is the superconducting Berry curvature while $\mathbf{a}$ labels bands of the Bogoliubov–de Gennes (BdG) Hamiltonian
\begin{align}\label{eq:H}
\mathbf H(\bm{k})=\left(\begin{array}{cc}
		H_0(\bm{k}) & \boldsymbol{\Delta}_{\bm k} \\
		\boldsymbol{\Delta}_{\bm k} & -H_0(\bm{k})
	\end{array}\right).
\end{align}
The BdG Hamiltonian \eqref{eq:H} is written in the Nambu basis $\{\psi_{ \bm k},\Lambda^\dagger\psi_{-\bm k}^\dagger\}$ where $\psi_{\bm k}$ is a collection of fermion operators in the band basis, and $\boldsymbol{\Delta}_{\bm k}^\dagger=\boldsymbol{\Delta}_{\bm k}$. The time reversal transformation works on $\psi_{\bm k}$ as $\psi_ {\bm k}\rightarrow \Lambda K \psi_{\bm k}$, where $K$ is complex conjugation and $\Lambda$ is some unitary operation.  

The energy eigenvalues of the BdG Hamiltonian are ordered into pairs $\pm \mathbf E_{1,\bm{k}},\pm \mathbf E_{2,\bm{k}},\cdots$ with $\mathbf E_{1,\bm k}<\mathbf E_{2,\bm k}<\cdots$ and $\mathbf a=\pm1,\pm2,\cdots$ are used to label the bands such that $\mathbf E_{-\mathbf a,\bm{k}}=-\mathbf E_{\mathbf a,\bm{k}}$. Assuming only the intraband pairing is present, we have $\mathbf E_{|\mathbf a|,\bm{k}}=(\epsilon_{|\mathbf a|,\bm{k}}^2+\varDelta_{|\mathbf a|,\bm{k}}^2)^{1/2}$, where $\epsilon_{a,\bm{k}}$ is the normal state band energy ($a=1,2,...$). We assume that only the $a=1$ band crosses the Fermi level, and all other bands lie strictly above or below it. This allows us to define the interband energy scale as $t\equiv \min_b \min_{\bm k} |\epsilon_{b,\bm{k}}|$ ($b\neq 1$). The gap function of the first band is defined as $\varDelta_{\bm{k}}\equiv \varDelta_{1,\bm{k}}$. We will consider the weak-pairing limit $|\varDelta_{\bm{k}}|\ll t$, so only the lowest-energy states with $\mathbf a=\pm 1$ need to be included to the leading-order $\varDelta_{\bm k}/t$ expansion of the Berry curvature
\begin{align}
	\boldsymbol \Omega ^{\mathbf a}_{\bm{k}}  \doteq -\mathop{\mathrm{Im}}\left\{\frac{\mathop{\mathrm{Tr}}[\mathbf P_{\mathbf a} \partial_{k_x} \mathbf H \mathbf P_{-\mathbf a} \partial_{k_y}\mathbf H \mathbf P_{\mathbf a}]}{(\mathbf E_{\mathbf a}-\mathbf E_{-\mathbf a})^2}-(x\leftrightarrow y)\right\},\label{eq:Omega}
\end{align}
where $\mathbf P_{\mathbf a}\equiv |\mathbf u^{\mathbf a}_{\bm{k}}\rangle\langle {\mathbf u}^{\mathbf a}_{\bm{k}}|$ is the projector in the Nambu space.

After a small phase twist $\varphi$ is turned on, $\mathbf H, \mathbf E_{\pm\mathbf a}, \mathbf P_{\pm \mathbf a}$ appearing in Eq. (\ref{eq:Omega}) all receive some corrections, which propagate through the THE formulas in Eqs. \eqref{eq:kappa} and \eqref{eq:sigma}. However, as shown in \cite{SM}, only the change in $\mathbf P_{\pm \mathbf a}$ is important when the unperturbed ground state obeys TRS. In the end, the intraband pairing contribution gives 
\begin{align}
	\delta \boldsymbol\Omega^{\mathbf a,\text{intra.}}_{\bm{k}}&\doteq \varphi\sum_{\mathbf b\neq \pm\mathbf a}4 \mbox{Im\,Tr}[ ( \partial_{k_x}\mathbf P_{\mathbf a} ) ( \partial_{k_y}\mathbf P_{-\mathbf a} ) \mathbf P_{\mathbf b} (\partial_\varphi\mathbf P_{\mathbf a} ) \mathbf P_{\mathbf a}]\notag\\
    &\qquad-(x\leftrightarrow y),
\end{align}
To the leading order of $\varDelta_{\bm k}/t$ expansion and focusing on the $\mathbf a = \mathbf 1$ band, we find
\begin{equation}
	\delta \boldsymbol\Omega^{\mathbf 1,\text{intra.}}_{\bm{k}}=-\varphi\cdot\frac{\varDelta_{\bm{k}}^2}{4\mathbf E_{\mathbf 1}^3}\big(v_x \partial_{k_y} - v_y \partial_{k_x}\big)\langle u^1_{\bm{k}}|\hat L|u^1_{\bm{k}}\rangle,\label{eq:Omega_intra}
\end{equation}
with $v_{x,y}\equiv \partial_{k_{x,y}} \epsilon_{1,\bm{k}}$ the normal state Fermi velocity. 

The change in Berry curvature is now fully described with normal state wave functions, and exhibits high concentration near the Fermi surface due to  $\mathbf{E}_{\mathbf 1}^3$ in the denominator. Identifying $\hat L|u_{\bm k}^1\rangle=4i\partial_\varphi|u_{\bm k}^1\rangle$, the derivative $\partial_{k_y}\langle u_{\bm k}^1|\hat L|u_{\bm k}^1\rangle=4\Omega_{y,\varphi}$ becomes the \emph{$\varphi$-twist Berry curvature}, with one component along the momentum direction and the other along the phase twist $\varphi$. Equation \eqref{eq:Omega_intra} thus demonstrates how the change of Berry curvature in the superconducting state is intricately composed of the gap function, and a mixed Berry curvature in the momentum-phase space. Plugging Eq. (\ref{eq:Omega_intra}) into Eq. (\ref{eq:kappa}) and after some efforts, the main result of our paper Eq. (\ref{eq:main_result}) is established.

\textit{Interband pairing and nodal superconductivity}.---
When interband pairing is present, the calculation becomes more sophisticated but the final result turns out to be simple. Introducing the normal state projector $P_c\equiv |u^c_{\bm{k}}\rangle\langle u^c_{\bm{k}}|$, the interband pairing gives $\boldsymbol\Delta_{\bm{k}}^{\text{inter.}}=\sum_{b\neq c}P_b \boldsymbol\Delta_{\bm{k}} P_c$, which can be eliminated from the BdG Hamiltonian in Eq. (\ref{eq:H}) by a small unitary rotation 
\begin{equation}
    e^{i\mathcal S\otimes\bm\tau_2}\mathbf H(\bm k) e^{-i\mathcal S\otimes\bm\tau_2}\doteq H_0(\bm k)\otimes\bm\tau_3 + \boldsymbol\Delta_{\bm k}^{\text{intra.}}\otimes\bm\tau_1 .
\end{equation}
Here $\bm\tau$ are Pauli matrices in the Nambu space, and $\mathcal S \equiv \sum_{a\neq b}P_a\boldsymbol\Delta_{\bm k}P_b/(\epsilon_{a,\bm k}-\epsilon_{b,\bm k})$ can be viewed small simply because $\boldsymbol\Delta_{\bm k}/(\epsilon_{a,\bm k}-\epsilon_{b,\bm k})\sim\varDelta_{\bm k}/t$ is small in the weak-pairing limit. Based on this useful property, all previous perturbative analysis for the intraband pairing can be extended to interband pairing as well, with only one modification replacing the projector $\mathbf P_{\mathbf a}$ by $\widetilde{\mathbf P}_{\mathbf a} = e^{-i\mathcal S\otimes\bm\tau_2}\mathbf P_{\mathbf a} e^{i\mathcal S\otimes\bm\tau_2}$. Collecting all $\varphi$-linear terms contributing to the change of the Berry curvature in Eq. \eqref{eq:Omega} seems to be more complicated, but to the leading order of $\varDelta_{\bm k}/t$
the interband contribution can be neatly arranged as \cite{SM}
\begin{align}
	&\delta \boldsymbol\Omega^{\mathbf 1,\text{inter.}}_{\bm{k}}=\varphi\cdot\frac{- {\mathbf d}_{\bm{k}}\cdot(\partial_{k_x}{\mathbf d}_{\bm{k}}\times \partial_{k_y} {\mathbf d}_{\bm{k}})}{2\mathbf E^3_\mathbf 1},\label{eq:Omega_inter} \\
    &{\mathbf d}_{\bm{k}}\equiv(\varDelta_{\bm{k}}, G_{\bm{k}},\epsilon_{1,\bm{k}}),\quad G_{\bm{k}}\equiv\frac{-1}{2}\mbox{Re}[\langle u^1_{\bm{k}}|\boldsymbol{\Delta}^{\text{inter.}}_{\bm{k}}\hat L|u^1_{\bm{k}}\rangle].\nonumber
\end{align}
Equation \eqref{eq:Omega_inter} is clearly reminiscent of the Berry curvature of the effective two-band model%
\begin{equation}\label{eq:H_eff}
	\mathbf H_{\text{eff}}=\varphi\cdot\partial_\varphi \epsilon_{1,\bm{k}}\boldsymbol\tau_0+\epsilon_{1,\bm{k}}\boldsymbol\tau_3+\varDelta_{\bm{k}}\boldsymbol\tau_1+\varphi\cdot G_{\bm{k}}\boldsymbol\tau_2.
\end{equation}
within the perturbative regime. Indeed, we show that $\mathbf H_{\text{eff}}$ is exactly the low-energy effective Hamiltonian of the TBS itself with insertion of phase twist \cite{SM} $\mathbf H_{\text{eff}}^{\mathbf a}=\widetilde{\mathbf P}_{\mathbf a}\mathbf H[\varphi]\widetilde{\mathbf P}_{\mathbf a}\doteq\widetilde{\mathbf P}_{\mathbf a}\mathbf H\widetilde{\mathbf P}_{\mathbf a}+\varphi\cdot\widetilde{\mathbf P}_{\mathbf a}(\partial_\varphi\mathbf H)\widetilde{\mathbf P}_{\mathbf a}$, so could serve as a faithful model to compute $\delta\mathbf\Omega_{\bm k}^{\mathbf 1,\text{inter.}}$.

In calculating $\delta \kappa_{xy}$ in Eq. \eqref{eq:kappa} one should add the two Berry curvature contributions $\delta \boldsymbol\Omega^{\mathbf 1}_{\bm{k}} = \delta \boldsymbol\Omega^{\mathbf 1,\text{intra.}}_{\bm{k}} + \delta \boldsymbol\Omega^{\mathbf 1,\text{inter.}}_{\bm{k}}$ in Eq. \eqref{eq:sigma}. Among the two, the intraband Berry curvature part can be converted to the Fermi surface integral form in Eq. (\ref{eq:main_result}) but not the interband part. For calculating the latter contribution we can directly use Eqs. \eqref{eq:kappa}-\eqref{eq:sigma} with $\delta \boldsymbol\Omega^{\mathbf 1,\text{inter.}}_{\bm{k}}$ given in Eq. \eqref{eq:Omega_inter}.

Although in principle the interband contribution for SATHE cannot be neglected, there are many physical situations in which the interband pairing and thus the interband contribution to THE does become small. For instance, if each monolayer of TBS itself is superconducting as in a vertical JJ, $\boldsymbol\Delta_{\bm k}$ in Eq. (\ref{eq:H}) is diagonal in the monolayer band basis. For small twist angle $\theta\ll1$, a unitary transformation to the band basis also induces interband pairing components in $\boldsymbol\Delta_{\bm k}$, which are proportional to $\theta$ and can be ignored. If the system is an intrinsic TBS like TBG, the intra- and inter-band pairings should be determined self-consistently. For example, for boson-mediated superconductivity, the multiband Eliashberg formulation may be applied, which is characterized by the electron-boson coupling matrix $[\alpha^2(\omega)F(\omega)]_{ab}$, where $a,b$ label bands. Either in the regime that intraband coupling is dominant $[\alpha^2(\omega)F(\omega)]_{a=b}\gg [\alpha^2(\omega)F(\omega)]_{a\neq b}$, or in the regime that the intraband coupling is dominant $[\alpha^2(\omega)F(\omega)]_{a=b}\ll [\alpha^2(\omega)F(\omega)]_{a\neq b}$, it is easy to show that only the intraband pairing is significant \cite{dolgov2009interband}.

However, there is one particular scenario in which the interband contributions may be dominant, and that is when the superconductivity in the TBS is nodal. In this case, the node would develop a mass gap $m_{\bm k}=\varphi\cdot G_{\bm{k}}$ right at the nodal point upon the application of a phase twist $\varphi$, and a Chern number transfer $\Delta C=\pm \frac{1}{2}$ between the low-energy BdG bands occurs per node. When the net transferred Chern numbers $n$ from all the nodes is nonzero, the system becomes a chiral topological superconductor with quantized THE $\delta\kappa_{xy}^{\text{inter.}}\equiv\varphi\cdot\chi_{\varphi xy}^{\text{inter.}}=ng_0$ in the low-temperature limit $k_BT\ll |m|$, where $\chi_{\varphi xy}^{\text{inter.}}$ is the interband transport coefficient for SATHE, and $g_0\equiv\pi k_B^2T/(6\hbar)$ is the quantum of thermal conductance. In our perturbative treatment, $\chi_{\varphi xy}^{\text{inter.}}$ diverges in the low-$T$ regime. This supercurrent-driven topological superconductivity has been discussed in the context of twisted cuprate bilayers \cite{volkov2023current,volkov2023magic,can2021high,song2022doping} and twisted NbSe$_2$ heterostructures \cite{hu2022engineering}.

\textit{Applications to FeSe and cuprates}.---
To demonstrate the possibility of observing SATHE in real materials we consider two examples: vertical JJ's formed by twisted nodeless FeSe, and the nodal cuprate superconducting films. The $\varDelta_{\bm{k}}$ in both of these 2D superconductors has been well characterized by ARPES due to the large energy scale of $\varDelta_{\bm{k}}$ and high transition temperatures. Note that due to the dimensionless nature of the SATHE response, the supercurrent-induced $\kappa_{xy}$ only depends on the ratio $\varDelta_{\bm{k}}/k_B T$ as in Eq. \eqref{eq:main_result} and the SATHE response of low-$T_c$ superconductors after appropriate rescaling must be similar to the examples we consider now.

\begin{figure}[!ht]
	\centering
	\includegraphics[scale=0.4]{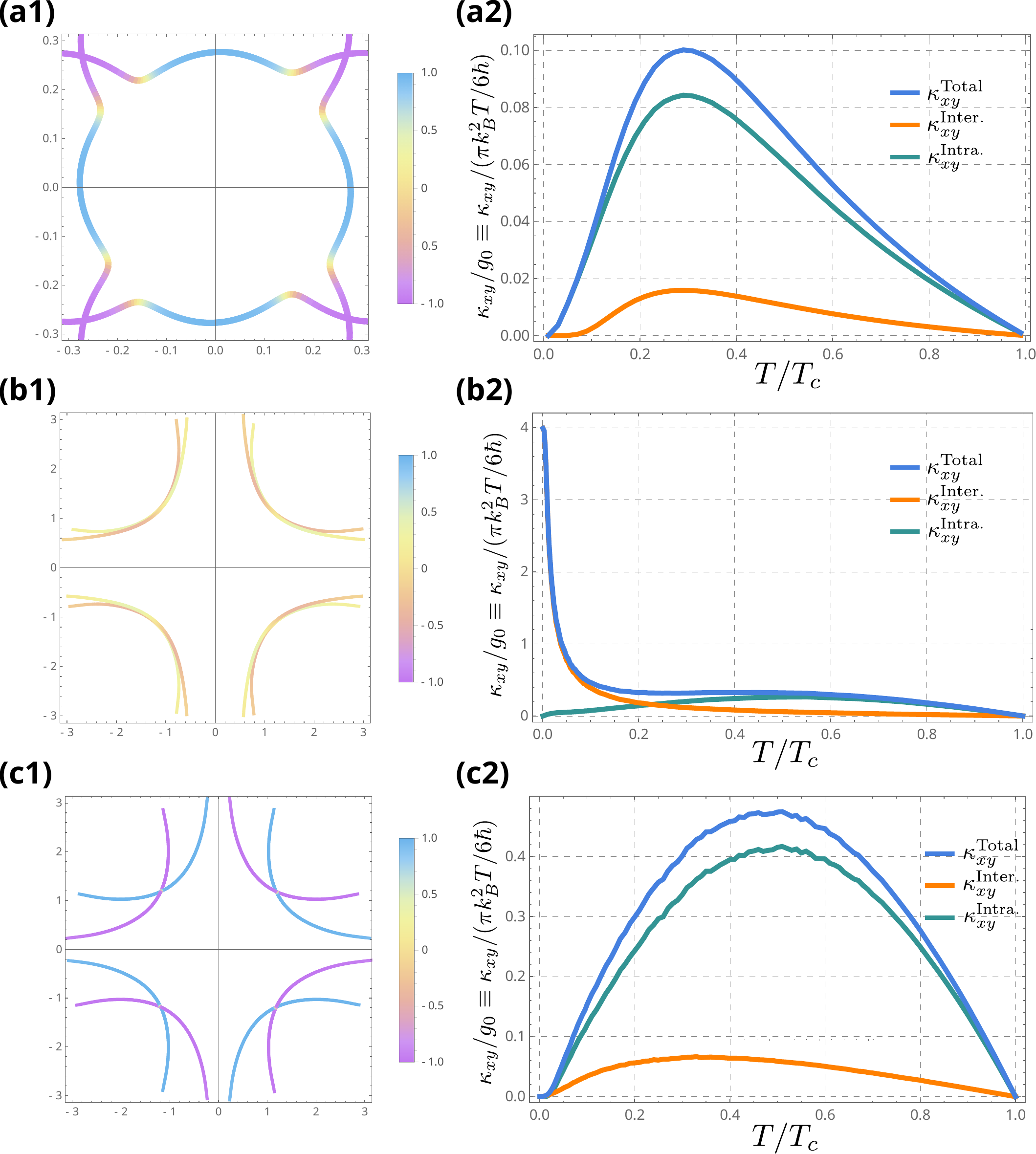}
	\caption{(a): JJ formed by twisted bilayer FeSe. (b) and (c): the topological non-trivial and trivial phases induced by the supercurrent in JJ formed by twisted bilayer cuprates. The twisted Fermi surfaces are shown on the left panels. For FeSe the moir\'e zone folding effect plays an important role on the topology of reconstructed Fermi surfaces. The color scheme on each Fermi surface represents the strength of $\langle u_{\bm{k}}|\hat{L}|u_{\bm{k}}\rangle\in[-1,1]$, serving as the indicator of the layer-component of states. For all three cases, $\kappa_{xy}$ is computed in unit of the thermal conductance quantum $g_0$ as a function of temperature, and a simple mean-field temperature-dependence of $\varDelta_{\bm{k}}(T)=\varDelta_{\bm{k}}(T=0)\sqrt{1-T/T_c}$ is implemented. For the topologically nontrivial case (b), gaps $\sim 0.2$meV are generated around superconducting nodes, leading to a Chern number $C=8$ topological superconductivity, while for the topological trivial case (c) the net Chern number is zero.}
	\label{fig: FS and kappa_xy}
\end{figure}

Monolayer FeSe is reported to host a significant nodeless gap anisotropy of the form $\varDelta_{\bm{k}}=\varDelta_0+\varDelta_2\cos2\theta_{\bm{k}}+\varDelta_4\cos4\theta_{\bm{k}}$, where $\varDelta_0=9.9$meV, $\varDelta_2=-1.4$meV, and $\varDelta_4=1.2$meV \cite{zhang2016superconducting}. The $C_4$-rotation-related elliptic $M$-pockets positioned at $(\pm\frac{\pi}{2},\pm\frac{\pi}{2})$ within the two-iron BZ \cite{brouet2012impact,borisenko2016direct,yi2015observation} can be described by the $\bm{k}\cdot\bm{p}$ expansion within $\mathrm{Fe}$'s $\{3d_{xz},3d_{yz}\}$ orbitals \cite{agterberg2017resilient,ran2009nodal} as $H_M=\left(\frac{1}{2m}(k_x^2+k_y^2)-\mu\right)+a k_xk_y\tau_z$, where $\tau_z$ is a Pauli matrix in the orbital space, $\mu=0.08$eV, $1/2m=1.4$eV$\cdot\text{\AA}^2$ and $a=0.6$eV$\cdot\text{\AA}^2$. Since all Fermi pockets are near the zone boundary, moir\'e zone folding effect comes into play, and Bistritzer-MacDonald model \cite{bistritzer2011moire} is used to construct the normal state Hamiltonian. For example, to the lowest truncation of the moir\'e BZ, two hopping proccesses corresponding to $\delta\bm{q}^{t,b}=\pm 2K_M\sin\frac{\theta}{2}\hat{k}_y$ are included for all four of the $M$-pockets (see Fig.2 in SM \cite{SM}). We set $T_\perp^{d_{xz}^t\text{-}d_{xz}^b}(\delta\bm{q}^{t,b})=T_\perp^{d_{yz}^t\text{-}d_{yz}^b}(\delta\bm{q}^{t,b})\simeq T_\perp=15$meV and ignore all interorbital hoppings. After turning on such simple $T_\perp$ with a twist angle $\theta=11.5^\circ$, the Fermi surfaces reconstruct within the moir\'e BZ (see Fig. \ref{fig: FS and kappa_xy} (a1)). 

We secondly consider a twisted bilayer of cuprates \footnote{Note that the currently available cuprate van der Waals materials is Bi2212, which is a bilayer of Cu-O planes. Consequently, twisted \emph{double} bilayer is more experimental relevant at present \cite{zhu2021presence,zhao2021emergent}. Our twisted bilayer model can viewed as a minimal illustration on SATHE for twisted nodal superconductors.} with a $d$-wave gap anisotropy $\varDelta_{\bm{k}}=\varDelta_N\cos2\theta_{\bm{k}}$ \cite{kanigel2007protected,kondo2015point} and the reported relation $8.5k_BT_c=2\varDelta_N$ \cite{kendziora1996superconducting,anzai2013relation}. The normal state bilayer Hamiltonian is constructed by first taking the tight-binding model in Ref. \cite{eschrig2003effect} as the monolayer Hamiltonian, and then obtaining the interlayer tunneling $T_\perp(\bm{k})=t_z\big(\frac{1}{4}(\cos k_x^t-\cos k_y^t)(\cos k_y^b-\cos k_y^b)+a_0\big)$ from the detailed orbital analysis in Refs. \cite{song2022doping,markiewicz2005one}. The constant $a_0=0.4$ is determined from DFT simulations \cite{markiewicz2005one}. We set (i) $t_z=0.025t_0$ ($t_0$=leading hopping strength within the $\mathrm{CuO}_2$ plane) with $\theta=0.6^\circ$ (chiral topological superconductor) and (ii) $t_z=0.01t_0$ with $\theta=17.2^\circ$ (topologically trivial superconductor) and plot the corresponding Fermi surfaces in Fig. \ref{fig: FS and kappa_xy} (b1) and (c1), respectively \cite{volkov2023current,volkov2023magic}. Here we neglect the moir\'e zone folding effect for several reasons. First, based on the two-center approximation \cite{bistritzer2011moire}, far away from the zone boundary the moir\'e zone folding effect may be neglected. Second, the moir\'e zone folding effect for the Fermi surfaces near the zone boundary does not modify the Fermi surface topology, leaving the results qualitatively unchanged.

The interband and intraband contributions to SATHE for twisted bilayers of FeSe and cuprates are plotted in Fig. \ref{fig: FS and kappa_xy} (a2)-(c2). In our linearized scheme, $\kappa_{xy}$ and $\chi_{\varphi xy}$ are related simply by $\kappa_{xy} = \varphi \cdot \chi_{\varphi xy}$. We plot $\kappa_{xy}$ rather than $\chi_{\varphi xy}$ because the latter apparently diverges as $\varphi \rightarrow 0$ while $\kappa_{xy}^{\text{inter.}}$ remains finite in a supercurrent-driven topological superconductor as shown in Fig. \ref{fig: FS and kappa_xy} (b2). For illustrative purposes, phase twist $\varphi=1$ rad is chosen for computation \footnote{Josephson phase twist $\varphi=1$ rad is taken, primarily to exhibit the strength of the transport coefficient $\chi_{\varphi xy}^{\text{intra.}}$ from the intraband contribution along with the quantized behavior of $\kappa_{xy}$ arising from the interband one. The former component contributes to the thermal Hall conductivity simply as $\kappa_{xy}^{\text{intra.}}=\varphi\cdot\chi_{\varphi xy}^{\text{intra.}}$, as along as we remain within the \emph{linear regime} of the sine current-phase relation, which is satisfied simply because $\sin1.0\doteq0.84\approx1.0$. Therefore, the choice $\varphi=1$ rad can still be considered small in our perturbative treatment, and with such choice the plot of $\kappa_{xy}$ also directly reflects the strength of the SATHE transport coefficient $\chi_{\varphi xy}$ that we are concerned with.}. The interband contribution is evaluted by using the integral formula of Eq. \eqref{eq:main_result}. For the interband contribution we rely on Eqs. \eqref{eq:kappa}-\eqref{eq:sigma} with the Berry curvature obtained in Eq. \eqref{eq:Omega_inter}. 

As shown in Fig. \ref{fig: FS and kappa_xy}, the intraband contribution plays a dominant role for SATHE in both twisted FeSe (a2) and the topological trivial case of twisted cuprates (c2), while interband contribution dominates in topological nontrivial case of twisted cuprates (b2), especially in the low-temperature regime. 
Note that the $\kappa_{xy}$ in the proposed SATHE can reach $\sim 10^{-1}$ of the thermal conductance quantum when the gap anisotropy is significant, e.g., in the FeSe example. We conclude that SATHE can be a sizable effect, detectable in the foreseeable future.

\textit{Conclusion and Discussion}.---
We propose the supercurrent-induced anomalous thermal Hall effect, SATHE, as a new probe to the in-plane gap anisotropy of bilayer superconductors. Different from the conventional thermal Hall effect, the ground state preserve the time reversal symmetry. A pair of probes --- a vertically applied supercurrent and a horizontal temperature gradient --- is applied to induce the in-plane nonlinear thermal Hall response. Being a thermal response, it works on 2D superconductors where usual electrical probes fail. Since no external magnetic field needs to be applied, SATHE avoids the complications of vortices in the mixed state and probes purely the quasiparticle dynamics in superconductors. 

Within the BdG framework we showed that SATHE is sensitive to the in-plane gap anisotropy in the twisted bilayer superconductor, and could be large enough to serve as a new experimental probe of gap structure for atomically thin superconducting 2D crystals including twisted bilayer graphene systems, for which experimental probes have been limited due to the low dimensionality. Motivated by the discovery of novel quantum states in van der Waals materials, the experimental community has recently made significant progress towards measuring thermal transports in low dimensional systems. For instance, graphene-based Johnson noise thermometry has been developed \cite{fong2012ultrasensitive,talanov2021high} and used to measure thermal transports in graphene, carbon nanotubes and $\alpha$-RuCl$_3$ \cite{fong2013measurement,waissman2022electronic}. We believe SATHE can eventually find its place as an effective probe of 2D twisted materials, in particular for low $T_c$ superconductors where SATHE can serve as a sensitive measure of the gap anisotropy. 

Our results can be straightforwardly generalized to multi-layer superconductors, by replacing the operator $\hat L$ with the generator of the multi-layer Doppler shift.

\vspace{1em}
\begin{acknowledgments}
Y.R. and X.H. acknowledge the support from National Science Foundation under Grant No. DMR-1712128. J.H.H. was supported by the National Research Foundation of Korea (NRF) grant funded by the Korea government (MSIT) (No. 2023R1A2C1002644). 
\end{acknowledgments}

\bibliography{SATHE_bib}

% \appendix
\pagebreak
\onecolumngrid % start with one column environment

%%%%%%%%%% Merge with supplemental materials %%%%%%%%%%
%%%%%%%%%% Prefix a "S" to all equations, figures, tables and reset the counter %%%%%%%%%%
\setcounter{equation}{0}
\setcounter{figure}{0}
\setcounter{table}{0}
\setcounter{page}{1}
\renewcommand{\theequation}{S\arabic{equation}}
\renewcommand{\thefigure}{S\arabic{figure}}
\renewcommand{\bibnumfmt}[1]{[S#1]}
\renewcommand{\citenumfont}[1]{S#1}

% \makeatletter
%%%%%%%%%% Prefix a "S" to all equations, figures, tables and reset the counter %%%%%%%%%%

\begin{center}
    \textbf{Supplemental Material for ``Supercurrent-induced Anomalous Thermal Hall Effect as a New Probe to Superconducting Gap Anisotropy''}\\[1.5em]
    Xiaodong Hu, Jung Hun Han, and Ying Ran
\end{center}

\section{The Doppler shift and the gauge transformation due to the supercurrent}
The Doppler shift effect due to supercurrent has been discussed in textbooks and carefully in a recent paper by Crowley and Fu \cite{crowley2022supercurrent}. Below we follow the notation in the Appendix B of Ref. \cite{crowley2022supercurrent}.

To model a superconducting state in the presence of a finite supercurrent with supercurrent velocity $\mathbf u $, it is convenient to consider a moving reference frame $S'$ which is moving at a velocity $\mathbf u$ relative to the lab frame $S$. In the moving frame $S'$, there is no supercurrent and the superconducting order parameter is spatially uniform. 

Microscopically, the Galilean transformation between the frame-$S$ and frame-$S'$ is implemented by the unitary:
\begin{align}
	U_{\mathbf u}&=e^{\frac{i}{\hbar}\mathbf u\cdot \mathbf g},& \mathbf g&=M\mathbf R-\mathbf K t,
\end{align} 
where $\mathbf K$ is the total momentum and $\mathbf R$ is the center of mass:
\begin{align}
	\mathbf K&=\int d\mathbf r\,c_{\mathbf r,\sigma}^\dagger (-i\hbar \nabla) c_{\mathbf r,\sigma},&\mathbf R&=\frac{1}{M}\int d\mathbf r\,c_{\mathbf r,\sigma}^\dagger c_{\mathbf r,\sigma}m \mathbf r.
\end{align}
Here $M$ is the total mass, and $\sigma$ labels electron's spin. Under Galilean transformation, in the real-space basis:
\begin{align}
	U^\dagger_{\mathbf u} c^\dagger_{\mathbf r,\sigma}U_{\mathbf u}=c^\dagger_{\mathbf r-\mathbf u t,\sigma}e^{\frac{-i}{\hbar}m\mathbf r\cdot \mathbf u+\frac{i}{\hbar}m u^2t/2}.
\end{align}
Namely, the electron state $c^\dagger_{\mathbf r,\sigma}$ in frame-$S$ is transformed into the electron state $c^\dagger_{\mathbf r-\mathbf u t,\sigma}$ in frame-$S'$. The \emph{spatially uniform} pairing field in frame-$S'$ then can be represented in the real-space basis:
\begin{align}
	\hat\Delta_{S'}=\int d\mathbf r_1 d\mathbf r_2 \Delta(\mathbf r_1-\mathbf r_2) c^\dagger_{\mathbf r_1-\mathbf u t,\sigma_1}\epsilon_{\sigma_1,\sigma_2} c^\dagger_{\mathbf r_2-\mathbf u t,\sigma_2},
\end{align}
where we assumed spin-singlet pairing for simplicity, and $\epsilon_{\sigma_1,\sigma_2}$ is the Levi-Civita symbol. Performing the inverse Galilean transformation, one finds the pairing field in frame-$S$ is:
\begin{align}
	\hat\Delta_{S}=U_{\mathbf u}\hat\Delta_{S'}U^\dagger_{\mathbf u}=\int d\mathbf r_1 d\mathbf r_2 e^{\frac{i}{\hbar}m(\mathbf r_1+\mathbf r_2)\cdot \mathbf u}\Delta(\mathbf r_1-\mathbf r_2) c^\dagger_{\mathbf r_1,\sigma_1}\epsilon_{\sigma_1,\sigma_2} c^\dagger_{\mathbf r_2,\sigma_2},
\end{align}

One concludes that in the lab frame-$S$, the Cooper pair carries a nonzero center-of-mass momentum $2m\mathbf u$ due to the supercurrent. The mean-field Hamiltonian in the frame-$S$ is given by:
\begin{align}
	H^{MF}_{S}=H_0+\hat\Delta_{S},
\end{align}
where $H_0$ is the normal state Hamiltonian. One may now perform a space-dependent and time-independent gauge transformation $\mathbf U$ (which is different from the Galilean transformation) to eliminate the spatial dependence in the pairing field $\hat\Delta_S$:
\begin{align}
	\mathbf U &= e^{\frac{-i}{\hbar} \int d\mathbf r\,c^\dagger_{\mathbf r,\sigma} c_{\mathbf r,\sigma}m\mathbf r\cdot \mathbf u },& \mathbf U c^\dagger_{\mathbf r,\sigma} \mathbf U^\dagger&= e^{\frac{-i}{\hbar} m\mathbf r\cdot \mathbf u} c^\dagger_{\mathbf r,\sigma}
\end{align}

$H^{MF}_{S}$ after this unitary becomes:
\begin{align}
	\tilde H^{MF}_{S}\equiv \mathbf U H^{MF}_{S} \mathbf U^\dagger= \mathbf U H_0 \mathbf U^\dagger+\int d\mathbf r_1 d\mathbf r_2\Delta(\mathbf r_1-\mathbf r_2) c^\dagger_{\mathbf r_1,\sigma_1}\epsilon_{\sigma_1,\sigma_2} c^\dagger_{\mathbf r_2,\sigma_2},
\end{align}

In $\tilde H^{MF}_{S}$, the pairing field restores the form as if supercurrent is absent, and the normal state Hamiltonian becomes Doppler-shifted due to the gauge transformation. If we denote the phase factor $e^{\frac{-i}{\hbar}m\mathbf r\cdot\mathbf u}\equiv e^{-i\varphi(\mathbf r)/2}$, the supercurrent density is:
\begin{align}
	\mathbf J_s=-e n_s \mathbf u=-e  \frac{\hbar n_s}{2m} \nabla \varphi(\mathbf r),\label{eq:current_relation}
\end{align}
where we introduced the electron's superfluid density $n_s$, and $\varphi (\mathbf r)$ can be identified as the phase of the pairing order parameter, which recovers a well-known result. Generally speaking, the superfluid density may be spatially dependent $n_s(\mathbf r)$. For a uniform supercurrent density, due to current conservation, this implies that the superfluid velocity $\mathbf u(\mathbf r)$ would be spatially dependent. Namely, $\mathbf u(\mathbf r)$, or $\nabla \varphi$, would be larger when $n_s(\mathbf r)$ is smaller. In this situation, the unitary $\mathbf U$ should be modified as:
\begin{align}
	\mathbf U &= e^{\frac{-i}{\hbar} \int d\mathbf r\,c^\dagger_{\mathbf r,\sigma} c_{\mathbf r,\sigma}m \int_{\mathbf 0}^{\mathbf r} d\mathbf r'\cdot \mathbf u(\mathbf r') },& \mathbf U c^\dagger_{\mathbf r,\sigma} \mathbf U^\dagger&= e^{\frac{-i}{\hbar} m\int_{\mathbf 0}^{\mathbf r} d \mathbf r'\cdot \mathbf u(\mathbf r')} c^\dagger_{\mathbf r,\sigma}\equiv e^{-i\varphi(\mathbf r)/2}c^\dagger_{\mathbf r,\sigma},
\end{align}
and one still has $\mathbf u(\mathbf r)=\frac{\hbar}{2m}\nabla \varphi(\mathbf r)$.

The bottom line is that no matter the system is a bilayer Josephson junction or an instrinsic bilayer superconductor, a finite vertical supercurrent state is always modeled by the gauge transformation $c^\dagger_{\mathbf r,\sigma}\rightarrow e^{-i\varphi(\mathbf r)/2} c^\dagger_{\mathbf r,\sigma}$ that transforms the normal state electronic structure $H_0\rightarrow \mathbf U H_0 \mathbf U^\dagger$ while leaving the pairing field spatially uniform (as if the supercurrent is absent), as we have done in Eq.(4) in the main text. Here $\varphi(\mathbf r)$ satisfies Eq.(\ref{eq:current_relation}), and is only depending on the vertical coordinate $z$: $\varphi(\mathbf r)=\varphi(z)$.

\begin{figure}\label{fig:schematic_phase}
	\includegraphics[width=0.4\textwidth]{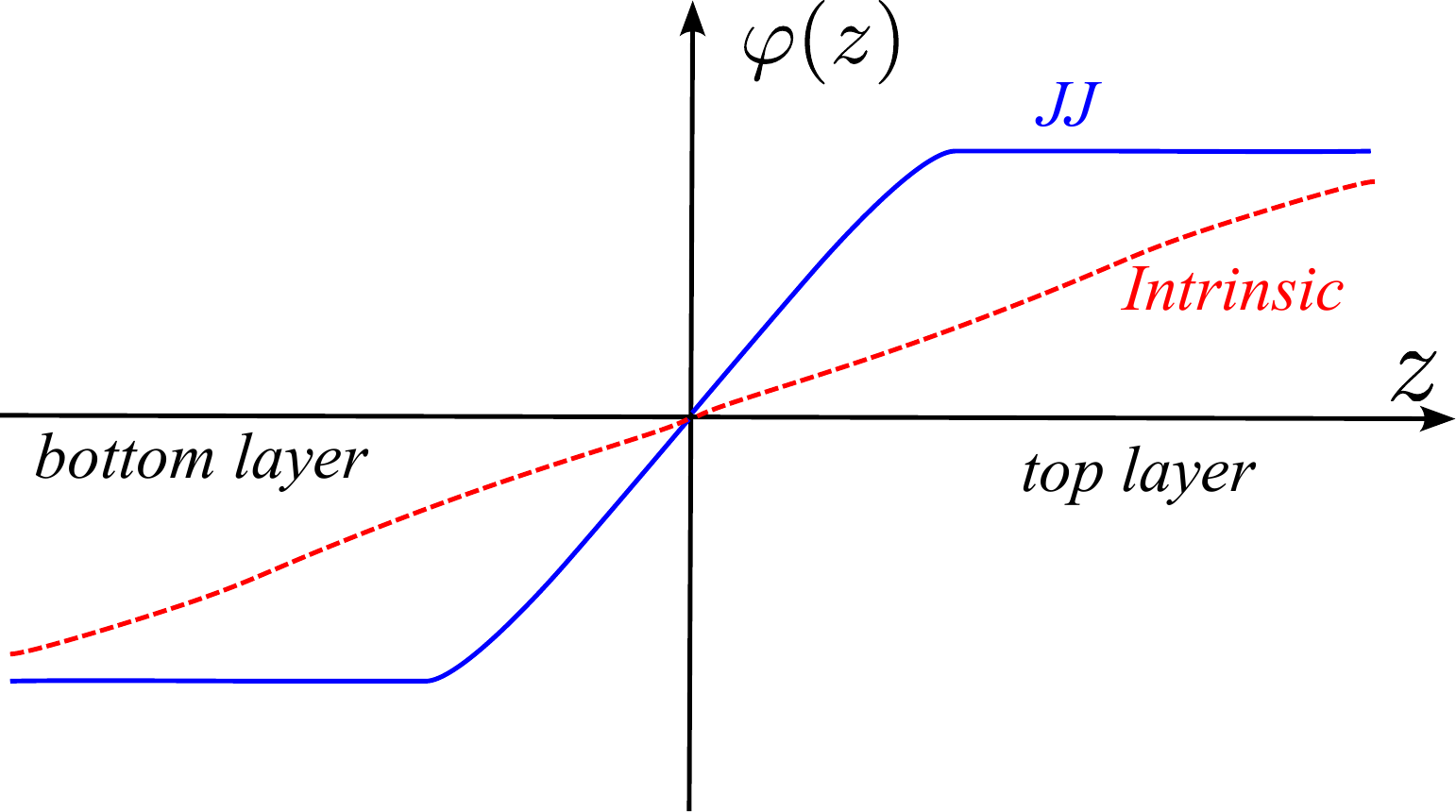}
	\caption{The schematic plot of the phase $\varphi$ involved in the gauge transformation to describe a finite supercurrent state. The two curves represent the case of a Josephson junction (JJ) and an intrinsic bilayer superconductor(Intrinisic) respectively.}
\end{figure}

The difference between the two cases lies in the details of the profile of $\varphi(z)$. In the case of a bilayer Josephson junction, the superfluid density is concentrated in each monolayers, while between the two layers $n_s$ is small. Consequently, the gradient of the phase $\varphi(z)$ is essentially located between the two layers. In the case of an intrinsic bilayer superconductor, $\varphi(z)$ would be a more smooth function of $z$. We schematically plot the phase $\varphi(z)$ of the gauge transformation for the two cases in Fig.\ref{fig:schematic_phase}.

We now consider the situation that $\varphi(z)$ interpolates the top layer $\delta\varphi/2$ to the bottom layer $-\delta\varphi/2$. The Doppler shift unitary is generated via $\hat L$: $\mathbf U=e^{-i\frac{\delta\varphi}{4} \hat L}$. 
\begin{itemize}
	\item In the case of a vertical JJ, $\hat L$ can written in the tight-binding basis, where the $\pm\mathbf 1$ blocks correspond to the top/bottom layers:
	\begin{align}
		\hat L_{\text{JJ}}=\hat c^\dagger \begin{pmatrix} \mathbf{1}&0\\0&\mathbf{-1}\end{pmatrix} c.
	\end{align} 
	\item In the case of an intrinsic bilayer superconductor, assuming the superfluid density is uniform, $\hat L$ is a diagonal matrix whose diagonal matrix elements are given by $\frac{z_\alpha}{z_0}$, where $z_\alpha$ is the atomic coordinate of the orbital-$\alpha$ and $\pm z_0$ is the position of the top/botom layers.
	\begin{align}
		\hat L_{\text{Intrinsic}}=\sum_\alpha c_\alpha^\dagger \frac{z_\alpha}{z_0} c_\alpha
	\end{align}
\end{itemize}

\section{Details on the Derivation of the Main Result} \label{app:Derivation_Details}

\subsection{Supercurrent-indcued Berry Curvature for Intraband Pairing}
Here we provide details of the derivation of the main result Eq.(1) in the main text. Because only $\mathbf a=\pm 1$ bands contribute to the low-energy Berry curvature, we will start with Eq.(8) in the main text.

Denoting the $c$-th eigenstate of $H_0(\bm{k})$ as $|u^c_{\bm{k}}\rangle$: $H_0(\bm{k})|u^c_{\bm{k}}\rangle=\epsilon_{c,\bm{k}}|u^c_{\bm{k}}\rangle$, and introducing the normal state projector $P_c\equiv |u^c_{\bm{k}}\rangle\langle u^c_{\bm{k}}|$, we can define the BdG projector into the low energy Hilbert space:
\begin{align}
	\mathbf P \equiv P_1\otimes \boldsymbol \tau_0,
\end{align}
where $\boldsymbol \tau_0$ is the identity Pauli matrix in the particle-hole space.
The low-energy BdG Hamiltonian is simply the two-by-two matrix, which perfectly decouples from the high energy Hilbert space if only intraband pairing is present (which we call \emph{intraband pairing assumption} below):
\begin{align}
	\mathbf P \mathbf H \mathbf P= P_1\otimes (\epsilon_{1,\bm{k}}\boldsymbol\tau_3+\varDelta_{\bm{k}}\boldsymbol\tau_1).
\end{align}
Here $\varDelta_{\bm{k}}$ is chosen to be real due to the time-reversal symmetry. In addition, Projector $\mathbf P$ satisfies an important identity with any operator of the following form:
\begin{align}
	\mathbf A^\vee\equiv A \otimes \boldsymbol\tau_0 \Rightarrow \mathbf P\mathbf A\mathbf P=\langle u^1_{\bm{k}}|A|u^1_{\bm{k}}\rangle\mathbf P,
\end{align}
which in turn indicates $\mathbf P_{-\mathbf a}\mathbf{A^\vee} \mathbf P_{\mathbf a}=0$.

After the pairing phase twist $\delta \varphi$ are turned on, $\mathbf H, \mathbf E_{\pm\mathbf a}, \mathbf P_{\pm \mathbf a}$ all receive perturbations in Eq.(8) in the main text. $\delta\mathbf H$ and $\delta\mathbf E_{\pm\mathbf a}$ do not contribute because $\partial_\varphi\mathbf H$ and $\partial_{k_x}\partial_\varphi\mathbf H$ both have the form of $\mathbf A^\vee$ \emph{under the intraband assumption}. For instance:
\begin{align}
	\partial_\varphi \mathbf H=\partial_\varphi H_0(\bm{k})\otimes\boldsymbol\tau_0\label{eq:partial_varphi}.
\end{align}
Thus we are left with the $\delta \mathbf P_{\pm \mathbf a}$ contribution only:
\begin{align}
\frac{\delta \boldsymbol \Omega ^{\mathbf a}}{\delta \varphi}\doteq \frac{- \mbox{Im\,Tr}[\mathbf P_{\mathbf a} \partial_{k_x} \mathbf H \mathbf P_{-\mathbf a} \partial_{k_y}\mathbf H \partial_\varphi\mathbf P_{\mathbf a}]}{\mathbf E_{\mathbf a}^2}-(x\leftrightarrow y).
\end{align}
Here we used the time-reversal symmetry combined with the particle-hole symmetry, which dictates that the $\delta \mathbf P_{-\mathbf a}$ contribution and $\delta \mathbf P_{\mathbf a}$ contribution are identical. After inserting $\mathbf 1=\sum_{\mathbf c} \mathbf{P}_{\mathbf c}$ in front of $\partial_\varphi \mathbf P_{\mathbf a}$, and noting that only terms with $\mathbf c\neq \pm \mathbf a$ contributes \emph{under the intraband pairing assumption}, the above expression can be rewritten as:
\begin{align}
	\frac{\delta \boldsymbol \Omega ^{\mathbf a}}{\delta \varphi}\doteq &\sum_{\mathbf b\neq \pm\mathbf a}\frac{\mbox{Im\,Tr}[\mathbf P_{\mathbf a} \partial_{k_x} \mathbf H \mathbf P_{-\mathbf a} \partial_{k_y}\mathbf H \mathbf P_{\mathbf b}\partial_\varphi\mathbf H \mathbf P_{\mathbf a}]}{\mathbf E_{\mathbf a}^2(\mathbf E_{\mathbf b}-\mathbf E_{\mathbf a})}-(x\leftrightarrow y)\notag\\
	=&\sum_{\mathbf b\neq \pm\mathbf a}\frac{ \mbox{Im\,Tr}[\mathbf P_{\mathbf a} \partial_{k_x} \mathbf H \mathbf P_{-\mathbf a} \partial_{k_y}\mathbf P_{\mathbf b}\partial_\varphi\mathbf H \mathbf P_{\mathbf a}]}{\mathbf E_{\mathbf a}^2(\mathbf E_{\mathbf b}-\mathbf E_{\mathbf a})}(\mathbf E_{\mathbf b}+\mathbf E_{\mathbf a})-(x\leftrightarrow y)\notag\\
	=&\sum_{\mathbf b\neq \pm\mathbf a}\frac{2 \mbox{Im\,Tr}[\mathbf P_{\mathbf a} \partial_{k_x} \mathbf H \mathbf P_{-\mathbf a} \partial_{k_y}\mathbf P_{\mathbf b}\partial_\varphi\mathbf H \mathbf P_{\mathbf a}]}{\mathbf E_{\mathbf a}(\mathbf E_{\mathbf b}-\mathbf E_{\mathbf a})}-(x\leftrightarrow y)\notag\\
	&\qquad+\sum_{\mathbf b\neq \pm\mathbf a}\frac{ \mbox{Im\,Tr}[\mathbf P_{\mathbf a} \partial_{k_x} \mathbf H \mathbf P_{-\mathbf a} \partial_{k_y}\mathbf P_{\mathbf b}\partial_\varphi\mathbf H \mathbf P_{\mathbf a}]}{\mathbf E_{\mathbf a}^2}-(x\leftrightarrow y).\label{eq:Omega_middle}
\end{align}
The term in the last line vanishes \emph{under the intraband pairing assumption}: $\mathbf P =\mathbf P_1+\mathbf P_{-1}$, $\sum_{\mathbf b\neq \pm\mathbf a}\partial_{k_y}\mathbf P_{\mathbf b}=-\partial_{k_y}\mathbf P$, and the fact that $\partial_{k_y}\mathbf P\partial_\varphi \mathbf H$ takes the form of $\mathbf A^\vee$. We therefore arrive at:
\begin{align}
	\frac{\delta \boldsymbol \Omega ^{\mathbf a,\text{intra.}}}{\delta \varphi}\doteq&\sum_{\mathbf b\neq \pm\mathbf a}4 \mbox{Im\,Tr}[\partial_{k_x}\mathbf P_{\mathbf a}  \partial_{k_y}\mathbf P_{-\mathbf a} \mathbf P_{\mathbf b}\partial_\varphi\mathbf P_{\mathbf a}\mathbf P_{\mathbf a}]-(x\leftrightarrow y)\notag\\
	=&\sum_{\mathbf b\neq \pm\mathbf a}\frac{2 \mbox{Im\,Tr}[\mathbf P_{\mathbf a} \partial_{k_x}\mathbf H \mathbf P_{-\mathbf a} \partial_{k_y}\mathbf H\mathbf P_{\mathbf b}\partial_\varphi\mathbf H\mathbf P_{\mathbf a}]}{ \mathbf E_{\mathbf a} (\mathbf E_{\mathbf b}-\mathbf E_{\mathbf a})(\mathbf E_{\mathbf b}-\mathbf E_{-\mathbf a})}-(x\leftrightarrow y)\notag\\
	\doteq &\sum_{b\neq 1}\frac{2 \mbox{Im\,Tr}[\mathbf P_{\mathbf a} \partial_{k_x}\mathbf H \mathbf P_{-\mathbf a} \partial_{k_y}\mathbf H(\mathbf P_{b}+\mathbf P_{-b})\partial_\varphi\mathbf H\mathbf P_{\mathbf a}]}{ \mathbf E_{\mathbf a} (\epsilon_{b}-\epsilon_{1})^2}-(x\leftrightarrow y)\notag\\
	= &\sum_{b\neq 1}\frac{2 \mbox{Im\,Tr}[\mathbf P_{\mathbf a} \partial_{k_x}\mathbf H \mathbf P_{-\mathbf a} \partial_{k_y}\mathbf H P_b\otimes\boldsymbol\tau_0 \partial_\varphi\mathbf H\mathbf P_{\mathbf a}]}{ \mathbf E_{\mathbf a} (\epsilon_{b}-\epsilon_{1})^2}-(x\leftrightarrow y)
\end{align}

Now we are ready to connect with the properties of the normal state. Introducing two-dimensional eigenvectors $|\boldsymbol\alpha_\mathbf c\rangle$ in the particle-hole subspace:
\begin{align}
	|\mathbf u^{\mathbf c}_{\bm{k}}\rangle=|u^{|\mathbf c|}_{\bm{k}}\rangle\otimes|\boldsymbol\alpha_\mathbf c \rangle,
\end{align}
and noting that
\begin{align}
	\partial_{k}\mathbf H=\partial_{\bm{k}} H_0(\bm{k})\otimes\boldsymbol\tau_3+O(\partial_{\bm{k}}\boldsymbol\Delta_{\bm{k}}),\label{eq:partial_k}
\end{align}
together with Eq.(\ref{eq:partial_varphi}), one has:
\begin{align}
	\frac{\delta \boldsymbol \Omega ^{\mathbf a,\text{intra.}}}{\delta \varphi}\doteq &\sum_{b\neq 1}\frac{2 |\langle \boldsymbol\alpha_\mathbf a|\tau_3| \boldsymbol\alpha_{-\mathbf a}\rangle|^2 \mbox{Im\,Tr}[P_1 \partial_{k_x}H_0 P_1 \partial_{k_y} H_0 P_b \partial_\varphi H_0 P_1]}{ \mathbf E_{\mathbf a} (\epsilon_{b}-\epsilon_{1})^2}-(x\leftrightarrow y)\notag\\
	=&\sum_{b\neq 1}\frac{2 |\langle \boldsymbol\alpha_\mathbf a|\tau_3| \boldsymbol\alpha_{-\mathbf a}\rangle|^2 v_x \mbox{Im\,Tr}[ P_1 \partial_{k_y} H_0 P_b \partial_\varphi H_0 P_1]}{ \mathbf E_{\mathbf a} (\epsilon_{b}-\epsilon_{1})^2}-(x\leftrightarrow y)\notag\\
	=&- \frac{v_x|\langle \boldsymbol\alpha_\mathbf a|\tau_3| \boldsymbol\alpha_{-\mathbf a}\rangle|^2     }{\mathbf E_{\mathbf a}}\Omega_{k_y,\varphi}-(x\leftrightarrow y)=- \frac{\varDelta_{\bm{k}}^2}{\mathbf E_{\mathbf a}^3}v_x  \Omega_{k_y,\varphi}-(x\leftrightarrow y),
\end{align}
where $v_x\equiv \partial_{k_x} \epsilon_{1,\bm{k}}$, and $\Omega_{k_y,\varphi}\equiv -2\mbox{Im}[\langle\partial_{k_y} u^1_{\bm{k}}|\partial_\varphi u^1_{\bm{k}}\rangle]$ is the \emph{normal-state} Berry curvature w.r.t. $k_y$ and $\varphi$. Using the fact that $|\partial_\varphi u^1_{\bm{k}}\rangle=\frac{-i}{4}\hat L |u^1_{\bm{k}}\rangle$, the normal-state Berry curvature $\Omega_{k,\varphi}$ can also be expressed as
\begin{align}
	\Omega_{k,\varphi}=\frac{1}{4}\partial_{\bm{k}}\langle u^1_{\bm{k}}|\hat L|u^1_{\bm{k}}\rangle,
\end{align}
so that Eq.(10) in the main text is obtained.

\subsection{Fermi Surface Integral}
The SATHE can be computed using Eq.(5) and Eq.(6) in the main text. Define
\begin{align}
	\boldsymbol\Omega(E)\equiv \sum_{\mathbf a}\int \frac{d^2k}{(2\pi)^2}\boldsymbol\Omega^\mathbf a_{\bm{k}}\delta(E-\mathbf E_{\mathbf a,\bm{k}}),
\end{align}
we have $\boldsymbol\sigma (E)=\boldsymbol\sigma(0)-\int_0^E dE'\boldsymbol\Omega(E')$. Here $\boldsymbol\sigma(0)=0$ due to the time-reversal symmetry and the nodeless assumption.

Because the change of the Berry curvature is concentrated near the Fermi surface due to the $\mathbf E_{\mathbf a}$ denominator, and proportional to $1/\varDelta_{\bm{k}}$ right at the Fermi surface, it is convenient to represent $\boldsymbol\Omega(E)$ as a Fermi surface integral as following:
\begin{align}
	\frac{\delta\boldsymbol\Omega(E)}{\delta\varphi}
	\doteq &\sum_{\text{FS}}\frac{1}{(2\pi)^2}\oint _{\text{FS},\epsilon_{\bm{k}}\equiv\pm\sqrt{E^2-\varDelta_{\bm{k}}^2}} dk_\parallel \frac{|E|}{|\epsilon_{\bm{k}}|}\frac{1}{\hbar |v_F|}\mathbf\Omega^{\mathop{\mathrm{sign}}(E)}_{\bm{k}}\notag\\
	\doteq &\sum_{\text{FS}}\frac{-2}{(2\pi)^2}\oint _{\text{FS}} dk_\parallel \frac{|E|}{\sqrt{E^2-\varDelta_{\bm{k}}^2}}\frac{1}{\hbar |v_F|}\frac{\varDelta_{\bm{k}}^2}{E^3}(v_x\Omega_{k_y,\varphi}-v_y\Omega_{k_x,\varphi})\notag\\
	=&\mathop{\mathrm{sign}}(E)\sum_{\text{FS}}\frac{-1}{2\pi^2}\oint _{\text{FS}} dk_\parallel \frac{\varDelta_{\bm{k}}^2}{E^2\sqrt{E^2-\varDelta_{\bm{k}}^2}}\frac{1}{\hbar |v_F|}(v_x\Omega_{k_y,\varphi}-v_y\Omega_{k_x,\varphi})\notag\\
	=&\mathop{\mathrm{sign}}(E)\sum_{\text{FS}}\frac{-\mathop{\mathrm{sign}}(v_F)}{2\pi^2\hbar}\oint _{\text{FS}} d\vec k_\parallel \cdot (\Omega_{k_x,\varphi},\Omega_{k_y,\varphi}) \frac{\varDelta_{\bm{k}}^2}{E^2\sqrt{E^2-\varDelta_{\bm{k}}^2}}\notag\\
	=&\mathop{\mathrm{sign}}(E)\sum_{\text{FS}}\frac{-\mathop{\mathrm{sign}}(v_F)}{8\pi^2\hbar}\oint _{\text{FS}} d k_\parallel \partial_{k_\parallel} \langle u_{\bm{k}}|\hat L|u_{\bm{k}} \rangle \frac{\varDelta_{\bm{k}}^2}{E^2\sqrt{E^2-\varDelta_{\bm{k}}^2}}.
\end{align}

Integrating out $E$ in advance, we get
\begin{align}
	\boldsymbol\sigma(E)=\sum_{\text{FS}}\frac{\mathop{\mathrm{sign}}(v_F)}{8\pi^2\hbar}\oint _{\text{FS}} d k_\parallel \partial_{k_\parallel} \langle u_{\bm{k}}|\hat L|u_{\bm{k}} \rangle \frac{\sqrt{E^2-\varDelta_{\bm{k}}^2}}{|E|}\Theta(|E|-|\varDelta_{\bm{k}}|).
\end{align}

We finally arrive at the result using Eq.(5) in the main text:
\begin{align}
	\frac{\delta\kappa_{xy}}{\delta\varphi T}=\sum_{\text{FS}}\frac{\mathop{\mathrm{sign}}(v_F)k_B^2}{16\pi^2\hbar}\oint _{\text{FS}} d k_\parallel \partial_{k_\parallel} \langle u_{\bm{k}}|\hat L|u_{\bm{k}} \rangle \xi\bigg(\frac{\varDelta_{\bm{k}}}{T}\bigg).
\end{align}

\section{Supercurrent-indcued Berry Curvature for Interband Pairing}
When interband pairing is present, additional terms need to be considered. However, one can always perform a small unitary rotation $e^{iS\otimes\boldsymbol\tau_2}$ to eliminate the interband pairing:
\begin{align}
	S\equiv \sum_{a\neq b}\frac{P_a\boldsymbol \Delta^{\text{inter.}}_{\bm{k}} P_b}{\epsilon_{a,\bm{k}}+\epsilon_{b,\bm{k}}}, \quad S^\dagger=S.
\end{align}
Since $\mathbf H = H_0\otimes\boldsymbol\tau_3+ (\boldsymbol \Delta^{\text{intra.}}+\boldsymbol\Delta^{\text{inter.}})\otimes \boldsymbol \tau_1$, we have:
\begin{align}
	e^{iS\otimes\boldsymbol\tau_2}\mathbf H e^{-iS\otimes\boldsymbol\tau_2}\doteq & H_0\otimes\boldsymbol\tau_3+(\boldsymbol \Delta^{\text{intra.}}+\boldsymbol\Delta^{\text{inter.}})\otimes \boldsymbol \tau_1+(i)[S\otimes\boldsymbol\tau_2,H_0\otimes\boldsymbol\tau_3]\notag\\
	=& H_0\otimes\boldsymbol\tau_3+(\boldsymbol \Delta^{\text{intra.}}+\boldsymbol\Delta^{\text{inter.}})\otimes \boldsymbol \tau_1+(i) (SH_0+H_0S)\otimes (i)\boldsymbol\tau_1\notag\\
	=&H_0\otimes\boldsymbol\tau_3+(\boldsymbol \Delta^{\text{intra.}}+\boldsymbol\Delta^{\text{inter.}})\otimes \boldsymbol \tau_1+(-1) \boldsymbol\Delta^{\text{inter.}}_{\bm{k}}\otimes \boldsymbol\tau_1=H_0\otimes\boldsymbol\tau_3+\boldsymbol \Delta^{\text{intra.}}\otimes \boldsymbol \tau_1
\end{align}

Denoting the projectors in the presence of interband pairing as $\widetilde {\mathbf P}_{\mathbf c}$, we have $\widetilde {\mathbf P}_{\mathbf c}\doteq e^{-i S\otimes\boldsymbol\tau_2}\mathbf P_{\mathbf c}e^{i S\otimes\boldsymbol\tau_2}$. Basically, in all the previous derivations for the intraband case, we only need to replace $\mathbf P_{\mathbf c}$ by $\widetilde {\mathbf P}_{\mathbf c}$, which corresponds to performing the small unitary transformation for the operators like $\partial_\varphi\mathbf H\rightarrow e^{i S\otimes\boldsymbol\tau_2}\partial_\varphi\mathbf H e^{-i S\otimes\boldsymbol\tau_2}$. In this way $\partial_\varphi \mathbf H$ has a correction $[i S, \partial_\varphi H_0]\otimes\boldsymbol\tau_2$ and is no longer $\propto \boldsymbol\tau_0$. We can leave all the energies $\mathbf E_{\mathbf c}$ unchanged since they receive second order contributions from $\boldsymbol\Delta^{\text{inter.}}$.

After inspection, to the leading order of $\varDelta/t$, one can identify three sources of interband contributions. First, we do need to consider $\delta_\varphi \partial_{\bm{k}} \mathbf H$ in Eq.(8) in the main text. Second, in the first line of Eq.(\ref{eq:Omega_middle}), $\mathbf b=-\mathbf a$ needs to be included. Third, the term in the last line of Eq.(\ref{eq:Omega_middle}) is no longer vanishing. We term them as \emph{Part-A,B,C} and compute them one by one below.

\emph{Part-A:} For the $\delta_\varphi \partial_{\bm{k}} \mathbf H$ contributions, 
\begin{align}
	\left.\frac{\delta\boldsymbol\Omega^{\mathbf a}_{\bm{k}}}{\delta\varphi}\right|_{A}=&\frac{-\mbox{Im\,Tr}\big[\mathbf P_{\mathbf a}\partial_{k_x}\mathbf H \mathbf P_{-\mathbf a} [iS\otimes\boldsymbol\tau_2,\partial_{k_y}\partial_{\varphi}\mathbf H] \mathbf P_{\mathbf a}\big]}{2\mathbf E^2_{\mathbf a}}-(x\leftrightarrow y)\notag\\
	=&\frac{-\mbox{Im\,Tr}\big[\mathbf P_{\mathbf a}(\partial_{k_x}H_0\otimes\boldsymbol\tau_3 +\partial_{k_x}\boldsymbol\Delta\otimes\boldsymbol\tau_1 ) \mathbf P_{-\mathbf a} [iS\otimes\boldsymbol\tau_2,\partial_{k_y}\partial_{\varphi}H_0\otimes\boldsymbol\tau_0] \mathbf P_{\mathbf a}\big]}{2\mathbf E^2_{\mathbf a}}-(x\leftrightarrow y)\notag\\
	=&-\langle u^1_{\bm{k}}|[iS,\partial_{k_y}\partial_{\varphi}H_0]|u^1_{\bm{k}}\rangle\notag\\
    &\qquad\times\frac{v_x\mbox{Im}\big[\langle\boldsymbol\alpha_{\mathbf a}|\boldsymbol\tau_3  |\boldsymbol\alpha_{-\mathbf a}\rangle\langle\boldsymbol\alpha_{-\mathbf a}|\boldsymbol\tau_2 |\boldsymbol\alpha_{\mathbf a}\rangle\big]+\langle u^1_{\bm{k}}|\partial_{k_x}\boldsymbol\Delta|u^1_{\bm{k}}\rangle \mbox{Im}\big[\langle\boldsymbol\alpha_{\mathbf a}|\boldsymbol\tau_1  |\boldsymbol\alpha_{-\mathbf a}\rangle\langle\boldsymbol\alpha_{-\mathbf a}|\boldsymbol\tau_2 |\boldsymbol\alpha_{\mathbf a}\rangle\big]}{2\mathbf E^2_{\mathbf a}}-(x\leftrightarrow y).
\end{align}
Noting a few identities:
\begin{align}
	&\boldsymbol\tau_2|\boldsymbol\alpha_{-\mathbf a}\rangle\langle\boldsymbol\alpha_{-\mathbf a}|\boldsymbol\tau_2=|\boldsymbol\alpha_{\mathbf a}\rangle\langle\boldsymbol\alpha_{\mathbf a}|\notag\\  
	&\partial_{k_x}\mathbf E_\mathbf a=\mbox{Tr}[\mathbf P_\mathbf a (\partial_{k_x} H_0\otimes \boldsymbol \tau_3+\partial_{k_x} \boldsymbol\Delta\otimes \boldsymbol \tau_1)\mathbf P_\mathbf a]=v_x \frac{\epsilon_{1,\bm{k}}}{\mathbf E_a}+\langle u^1_{\bm{k}}|\partial_{k_x}\boldsymbol\Delta|u^1_{\bm{k}}\rangle \frac{\varDelta_{\bm{k}}}{\mathbf E_\mathbf a}\notag\\
	&\partial_{k_x} \mathbf E_\mathbf a=\frac{\varDelta_{\bm{k}}\partial_{k_x}\varDelta_{\bm{k}}+\epsilon_{1,\bm{k}}v_x}{\mathbf E_\mathbf a}\notag\\
	&\Rightarrow \langle u^1_{\bm{k}}|\partial_{k_x}\boldsymbol\Delta|u^1_{\bm{k}}\rangle = \partial_{k_x}\varDelta_{\bm{k}}\notag\\
	&\langle\boldsymbol\alpha_{\mathbf a}|\boldsymbol\tau_1  |\boldsymbol\alpha_{\mathbf a}\rangle=\frac{\varDelta_{\bm{k}}}{\mathbf E_\mathbf a}, \;\;\langle\boldsymbol\alpha_{\mathbf a}|\boldsymbol\tau_3  |\boldsymbol\alpha_{\mathbf a}\rangle=\frac{\epsilon_{1,\bm{k}}}{\mathbf E_\mathbf a}
\end{align}
we have:
\begin{align}
	\left.\frac{\delta\boldsymbol\Omega^{\mathbf a}_{\bm{k}}}{\delta\varphi}\right|_{A}
	=&-\frac{\langle u^1_{\bm{k}}|[iS,\partial_{k_y}\partial_{\varphi}H_0]|u^1_{\bm{k}}\rangle}{2\mathbf E^3_{\mathbf a}}\big(-\varDelta_{\bm{k}} v_x+\epsilon_{1,\bm{k}}\partial_{k_x}\varDelta_{\bm{k}}\big)-(x\leftrightarrow y)\notag\\
\end{align}

\emph{Part-B:} This term is:
\begin{equation}
	\frac{\mbox{Im\,Tr}[\mathbf P_{\mathbf a} \partial_{k_x} \mathbf H \mathbf P_{-\mathbf a} \partial_{k_y}\mathbf H \mathbf P_{-\mathbf a}\partial_\varphi\mathbf H \mathbf P_{\mathbf a}]}{\mathbf E_{\mathbf a}^2(\mathbf E_{-\mathbf a}-\mathbf E_{\mathbf a})}-(x\leftrightarrow y)
\end{equation}
Note that $\mathbf P_{-\mathbf a} \partial_{k_y}\mathbf H \mathbf P_{-\mathbf a}=\partial_{k_y}\mathbf E_{\mathbf -a}\mathbf P_{-\mathbf a}$,
\begin{align}
	\left.\frac{\delta\boldsymbol\Omega^{\mathbf a}_{\bm{k}}}{\delta\varphi}\right|_{B}=&\frac{\partial_{k_y}\mathbf E_{\mathbf a}}{2\mathbf E_{\mathbf a}^3}\mbox{Im\,Tr}[\mathbf P_{\mathbf a} (\partial_{k_x}H_0\otimes\boldsymbol\tau_3+\partial_{k_x}\boldsymbol\Delta\otimes\boldsymbol\tau_1) \mathbf P_{-\mathbf a} [iS,\partial_\varphi H_0]\otimes\boldsymbol\tau_2 \mathbf P_{\mathbf a}]-(x\leftrightarrow y)\notag\\
	=&\frac{\partial_{k_y}\mathbf E_{\mathbf a}}{2\mathbf E_{\mathbf a}^3}\langle u^1_{\bm{k}}|[iS,\partial_\varphi H_0]|u^1_{\bm{k}}\rangle\mbox{Im\,Tr}[\mathbf P_{\mathbf a} (\partial_{k_x}H_0\otimes\boldsymbol\tau_3+\partial_{k_x}\boldsymbol\Delta\otimes\boldsymbol\tau_1) \mathbf P_{-\mathbf a}\boldsymbol\tau_2 \mathbf P_{\mathbf a}]  -(x\leftrightarrow y)\notag\\
	=&\frac{\partial_{k_y}\mathbf E_{\mathbf a}}{2\mathbf E_{\mathbf a}^3}\langle u^1_{\bm{k}}|[iS,\partial_\varphi H_0]|u^1_{\bm{k}}\rangle\mbox{Im\,Tr}[\mathbf P_{\mathbf a} (\partial_{k_x}H_0\otimes\boldsymbol\tau_3+\partial_{k_x}\boldsymbol\Delta\otimes\boldsymbol\tau_1) \boldsymbol\tau_2 \mathbf P_{\mathbf a}]  -(x\leftrightarrow y)\notag\\
	=&\frac{\partial_{k_y}\mathbf E_{\mathbf a}}{2\mathbf E_{\mathbf a}^3}\langle u^1_{\bm{k}}|[iS,\partial_\varphi H_0]|u^1_{\bm{k}}\rangle (-v_x\langle\boldsymbol\alpha_{\mathbf a}|\boldsymbol\tau_1| \boldsymbol\alpha_{\mathbf a}\rangle + \langle u^1_{\bm{k}}|\partial_{k_x}\boldsymbol\Delta|u^1_{\bm{k}}\rangle \langle\boldsymbol\alpha_{\mathbf a}|\boldsymbol\tau_3| \boldsymbol\alpha_{\mathbf a}\rangle ) -(x\leftrightarrow y)\notag\\
	=&\frac{\partial_{k_y}\mathbf E_{\mathbf a}}{2\mathbf E_{\mathbf a}^4}\langle u^1_{\bm{k}}|[iS,\partial_\varphi H_0]|u^1_{\bm{k}}\rangle \big(-\varDelta_{\bm{k}} v_x+\epsilon_{1,\bm{k}}\partial_{k_x}\varDelta_{\bm{k}}\big) -(x\leftrightarrow y)\notag\\
	=&\frac{(v_y\epsilon_{1,\bm{k}}+\varDelta_{\bm{k}}\partial_{k_y}\varDelta_{\bm{k}})}{2\mathbf E_{\mathbf a}^5}\langle u^1_{\bm{k}}|[iS,\partial_\varphi H_0]|u^1_{\bm{k}}\rangle \big(-\varDelta_{\bm{k}} v_x+\epsilon_{1,\bm{k}}\partial_{k_x}\varDelta_{\bm{k}}\big) -(x\leftrightarrow y)
\end{align}
After $(x\leftrightarrow y)$ antisymmetrization, we get
\begin{align}
	\left.\frac{\delta\boldsymbol\Omega^{\mathbf a}_{\bm{k}}}{\delta\varphi}\right|_{B}=&\frac{-1}{2\mathbf E_{\mathbf a}^3}\langle u^1_{\bm{k}}|[iS,\partial_\varphi H_0]|u^1_{\bm{k}}\rangle v_x\partial_{k_y}\varDelta_{\bm{k}} -(x\leftrightarrow y)
\end{align}

\emph{Part-C:} Note that 
\begin{align}
\partial_{k_y} \widetilde{\mathbf P}_{\mathbf b}\doteq \partial_{k_y}\big[\mathbf P_{\mathbf b}+[-i S\otimes\boldsymbol\tau_2,\mathbf P_{\mathbf b}]\big]\doteq e^{-i S\otimes\boldsymbol\tau_2}\partial_{k_y}\mathbf P_{\mathbf b}e^{i S\otimes\boldsymbol\tau_2}+[-i \partial_{k_y}S\otimes\boldsymbol\tau_2,\mathbf P_{\mathbf b}]
\end{align}
The contribution from the last line of Eq.(\ref{eq:Omega_middle}) becomes:
\begin{align}
	\left.\frac{\delta\boldsymbol\Omega^{\mathbf a}_{\bm{k}}}{\delta\varphi}\right|_{C}\doteq &\frac{ -\mbox{Im\,Tr}[\mathbf P_{\mathbf a} \partial_{k_x} \mathbf H \mathbf P_{-\mathbf a} \partial_{k_y}\mathbf P [iS\otimes \boldsymbol\tau_2,\partial_\varphi\mathbf H] \mathbf P_{\mathbf a}+\mathbf P_{\mathbf a} \partial_{k_x} \mathbf H \mathbf P_{-\mathbf a} [-i \partial_{k_y}S\otimes\boldsymbol\tau_2,\mathbf P]\partial_\varphi\mathbf H \mathbf P_{\mathbf a}]}{\mathbf E_{\mathbf a}^2}-(x\leftrightarrow y)\notag\\
	=&\frac{-1}{\mathbf E_{\mathbf a}^3}\big(-\varDelta_{\bm{k}} v_x+\epsilon_{1,\bm{k}}\partial_{k_x}\varDelta_{\bm{k}}\big)\mbox{Re\,Tr}[\partial_{k_y}P_1[iS,\partial_\varphi H_0]P_1+P_1 i\partial_{k_y}S \partial_\varphi H_0 P_1]-(x\leftrightarrow y)\notag\\
	=&\frac{-1}{2\mathbf E_{\mathbf a}^3}\big(-\varDelta_{\bm{k}} v_x+\epsilon_{1,\bm{k}}\partial_{k_x}\varDelta_{\bm{k}}\big)\mbox{Re\,Tr}[2\partial_{k_y}P_1[iS,\partial_\varphi H_0]P_1+P_1 [i\partial_{k_y}S ,\partial_\varphi H_0] P_1]-(x\leftrightarrow y)
\end{align}

Let's add part-A and part-C together. 
\begin{align}
	\left.\frac{\delta\boldsymbol\Omega^{\mathbf a}_{\bm{k}}}{\delta\varphi}\right|_{A+C}=&\frac{-1}{2\mathbf E_{\mathbf a}^3}\big(-\varDelta_{\bm{k}} v_x+\epsilon_{1,\bm{k}}\partial_{k_x}\varDelta_{\bm{k}}\big)\partial_{k_y}\Big[\mbox{Re\,Tr}[P_1[iS,\partial_\varphi H_0]P_1]\Big]-(x\leftrightarrow y)\notag\\
	=&\frac{-1}{2\mathbf E_{\mathbf a}^3}\big(-\varDelta_{\bm{k}} v_x+\epsilon_{1,\bm{k}}\partial_{k_x}\varDelta_{\bm{k}}\big)\partial_{k_y}\langle u^1_{\bm{k}}|[iS,\partial_\varphi H_0]|u^1_{\bm{k}}\rangle-(x\leftrightarrow y)
\end{align}
The quantity $\langle u^1_{\bm{k}}|[iS,\partial_\varphi H_0]|u^1_{\bm{k}}\rangle$ here and in part-B can be easily computed, which we denote as $G_{\bm{k}}$:
\begin{align}
	G_{\bm{k}}\equiv&\langle u^1_{\bm{k}}|[iS,\partial_\varphi H_0]|u^1_{\bm{k}}\rangle=-2\mbox{Im}[\langle u^1_{\bm{k}}| S \partial_\varphi H_0|u^1_{\bm{k}}\rangle]=-2\sum_{b\neq 1}\mbox{Im}[\langle u^1_{\bm{k}}| S |u^b_{\bm{k}}\rangle\langle u^b_{\bm{k}}|\partial_\varphi H_0|u^1_{\bm{k}}\rangle]\notag\\
	&=-2\sum_{b\neq 1}\mbox{Im}\Big[\frac{\langle u^1_{\bm{k}}| \boldsymbol\Delta^{\text{inter.}} |u^b_{\bm{k}}\rangle\langle u^b_{\bm{k}}|\partial_\varphi H_0|u^1_{\bm{k}}\rangle}{\epsilon_{1,\bm{k}}+\epsilon_{b,\bm{k}}}\Big]\doteq-2\sum_{b\neq 1}\mbox{Im}\Big[\frac{\langle u^1_{\bm{k}}| \boldsymbol\Delta^{\text{inter.}} |u^b_{\bm{k}}\rangle\langle u^b_{\bm{k}}|\partial_\varphi H_0|u^1_{\bm{k}}\rangle}{-\epsilon_{1,\bm{k}}+\epsilon_{b,\bm{k}}}\Big]\notag\\
	&=2\sum_{b\neq 1}\mbox{Im}[\langle u^1_{\bm{k}}| \boldsymbol\Delta^{\text{inter.}} |u^b_{\bm{k}}\rangle\langle u^b_{\bm{k}} |\partial_\varphi u^1_{\bm{k}}\rangle]=2\mbox{Im}[\langle u^1_{\bm{k}}| \boldsymbol\Delta^{\text{inter.}} |\partial_\varphi u^1_{\bm{k}}\rangle]=\frac{-1}{2}\mbox{Re}[\langle u^1_{\bm{k}}| \boldsymbol\Delta^{\text{inter.}}\hat L | u^1_{\bm{k}}\rangle]
\end{align}
Putting together, we finally get
\begin{align}
	\frac{\delta\boldsymbol\Omega^{\mathbf a,\text{inter.}}_{\bm{k}}}{\delta\varphi}=\left.\frac{\delta\boldsymbol\Omega^{\mathbf a}_{\bm{k}}}{\delta\varphi}\right|_{A+B+C}=&\frac{-1}{2\mathbf E_{\mathbf a}^3}\big(-\varDelta_{\bm{k}} v_x+\epsilon_{1,\bm{k}}\partial_{k_x}\varDelta_{\bm{k}}\big)\partial_{k_y}G_{\bm{k}}+\frac{-1}{2\mathbf E_{\mathbf a}^3}G_{\bm{k}} v_x\partial_{k_y}\varDelta_{\bm{k}}-(x\leftrightarrow y)
\end{align}

It is also instructive to study the behavior of $\partial_\varphi\mathbf H$ in the low energy subspace:
\begin{align}
	\widetilde{\mathbf P} \partial_\varphi\mathbf H \widetilde{\mathbf P}=\mathbf P\partial_\varphi\mathbf H \mathbf P+\mathbf P[iS\otimes\boldsymbol\tau_2,\partial_\varphi\mathbf H]\mathbf P=\langle u^1_{\bm{k}}|\partial_\varphi H_0| u^1_{\bm{k}}\rangle\mathbf P +\langle u^1_{\bm{k}}|[iS,\partial_\varphi H_0] |u^1_{\bm{k}}\rangle\mathbf P\boldsymbol\tau_2\mathbf P
\end{align}
The low energy effective Hamiltonian in the presence of $\delta\varphi$ becomes:
\begin{align}
	\mathbf H_{\text{eff}}=\delta\varphi\partial_\varphi\epsilon_{1,\bm{k}}\boldsymbol\tau_0+\epsilon_{1,\bm{k}}\boldsymbol\tau_3+\varDelta_{\bm{k}}\boldsymbol\tau_1+\delta\varphi G_{\bm{k}}\boldsymbol\tau_2=\delta\varphi\partial_\varphi\epsilon_{1,\bm{k}}\boldsymbol\tau_0+\mathbf{d}_{\bm{k}}\cdot\vec{\boldsymbol{\tau}},
\end{align}
where we introduced vector $\mathbf{d}_{\bm{k}}\equiv (\varDelta_{\bm{k}},\delta\varphi G_{\bm{k}},\epsilon_{1,\bm{k}})$. We merely showed that the interband contribution can be faithfully computed using this effective 2-by-2 Hamiltonian:
\begin{align}
	\delta\boldsymbol\Omega^{\mathbf a,\text{inter.}}_{\bm{k}}=-2\mbox{Im\,Tr}[p_{\mathbf a}\partial_{k_x}p_{\mathbf a}\partial_{k_y}p_{\mathbf a}]=\frac{-d_{\bm{k}}\cdot(\partial_{k_x}\mathbf{d}_{\bm{k}} \times \partial_{k_y}\mathbf{d}_{\bm{k}})}{2\mathbf a|\mathbf{d}_{\bm{k}}|^3}\doteq \frac{-\mathbf{d}_{\bm{k}}\cdot(\partial_{k_x}\mathbf{d}_{\bm{k}} \times \partial_{k_y}\mathbf{d}_{\bm{k}})}{2\mathbf E_{\mathbf a}^3}\propto \delta\varphi
\end{align}
$p_{\mathbf a}=\frac{1}{2}(1+\mathbf a \frac{\mathbf{d}_{\bm{k}}\cdot\vec{\boldsymbol\tau}}{|\mathbf{d}_{\bm{k}}|})$ is the projector in this effective model.

When the superconductivity is nodal, near a node, the effective theory becomes a Dirac equation:
\begin{align}
\mathbf H^{\text{node}}_{\text{eff}}=\delta\varphi\partial_\varphi\epsilon_{1,\bm{k}}\boldsymbol\tau_0+\hbar v_F k_\perp\boldsymbol\tau_3+\hbar v_\Delta k_\parallel\boldsymbol\tau_1+m\boldsymbol\tau_2,
\end{align}
where $\hbar v_\Delta =\frac{\partial \varDelta_{\bm{k}}}{\partial k_\parallel}$, and $m=\delta\varphi G_{\bm{k}_{\text{node}}}$ is the mass gap generated by the supercurrent. It is easy to show that the nodal contribution to $\boldsymbol\sigma(E)$ is
\begin{align}
\boldsymbol\sigma^{\text{node}}(E)=\begin{cases}
      \frac{C}{2\pi}, & \text{if } -|m|<E<|m|, \\
      \frac{C |m|}{2\pi|E|}, & \text{otherwise}.
    \end{cases}
\end{align}
Here $C=\pm \frac{1}{2}$ is the Chern number transferred due to $m$. Performing the energy integral in Eq.(5) in the main text, in the low temperature limit $k_BT\ll |m|$ one recovers the quantized $\kappa^{\text{node}}_{xy}/T=\frac{k_B^2}{\hbar}\frac{C\pi}{12}$. However, in the high-temperature limit $m\ll k_B T$, we have:
\begin{align}
\frac{\kappa_{xy}^{\text{node}}}{T}\doteq  -\frac{k_B^2}{\hbar}\frac{C |m|}{2\pi k_BT}\int^{\infty}_{0} x f'(x) dx=\frac{k_B^2}{\hbar}\frac{C \ln 2 }{2\pi}\frac{ |m|  }{k_B T}.
\end{align}
Namely there is a $1/T$ tail in $\frac{\kappa_{xy}}{T}$.

As a final remark, in the absence of $\delta\varphi$, it is easy to show that the Berry curvature is nonsingular near the Fermi surface. Consequently, one does not need to consider the contribution to SATHE from $\partial_\varphi \mathbf E_{\mathbf a}$ in the leading order of $\varDelta/t$ expansion.

\section{Fermi Surface Reconstruction for Twisted Bilayer FeSe}
The effective $\bm{k\cdot p}$ model in the main text is written in the oribtal space of $\{3d_{xz},3d_{yz}\}$. Because the dominant hopping processes occur within \emph{the same} orbitals, we can treat each orbital separately, or equivalently go back to the single-iron Brillouin zone to work with one ellipse on each direction. The right horizontal elliptic $M$-pocket of monolayer FeSe is simply described with
\begin{equation}\label{H_mono_FeSe}
	h=-\widetilde{\mu} + a(k_x-k_M)^2 + bk_y^2
\end{equation}
with $\mu=-0.08$eV, $a=1.08\text{eV}\cdot\text{\AA}^2$ and $b=1.6\text{eV}\cdot\text{\AA}^2$. Denoting the rotated monolayer Hamiltonian as $h_{\pm\theta/2}(\bm{k})$, the rotated bilayer Hamiltonian \emph{without} interlayer tunnelings is then simply a two-by-two diagonal matrix $H_{\text{w/o }T_\perp}=\mathop{\mathrm{diag}}\{h_{\theta/2}(\bm{k}), h_{-\theta/2}(\bm{k})\}$.

Under small twisting angles, two-center approximation applies \cite{bistritzer2011moire} and the general tunneling strength between the top/bottom Bloch states $|\psi^t_{\alpha,\bm{k}^t}\rangle$ and $|\psi^b_{\beta,\bm{k}^b}\rangle$ takes the form of
\begin{equation}\label{two-center approximation}
	t_{\bm{k}^t,\bm{k}^b}^{\alpha\beta}\equiv\langle\psi_\alpha^t|H|\psi_\beta\rangle=\dfrac{1}{V}\sum_{\bm{G}^t,\bm{G}^b}\delta_{\bm{k}^t+\bm{G}^t,\bm{k}^b+\bm{G}^b}\cdot e^{-i\bm{G}^b\cdot\bm{\tau}^b_\alpha}\cdot t_{\alpha\beta}(\bm{k}^t+\bm{G}^t)\cdot e^{i\bm{G}^t\cdot\bm{\tau}^t_\beta},
\end{equation}
where $\bm{G}^{t,b}$ are the reciprocal vector of the top/bottom layer and $\bm{\tau}_\alpha^t$ and $\bm{\tau}_\beta^b$ are the sublattice position of the wannier centers of top/bottom states. Now that only one state is left in the single-iron Brillouin zone, we can simply take $t_{\alpha\beta}(\bm{k}^t+\bm{G}^t)=\delta_{\alpha\beta}t(\bm{k}^t+\bm{G}^t)$.

\begin{figure}[!htp]
	\centering
	\includegraphics[scale=0.4]{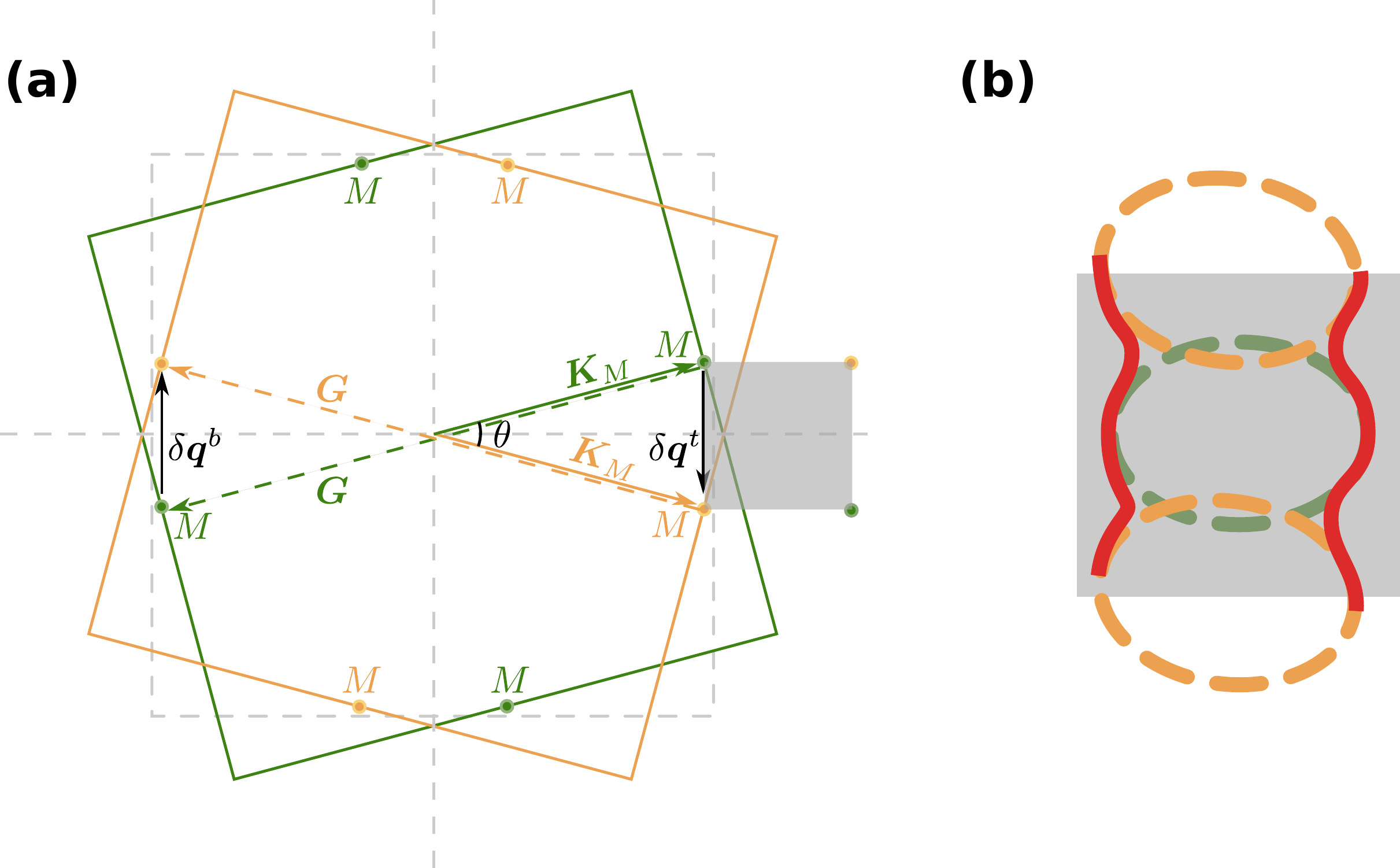}
	\caption{(a) Rotated top layer (green) and bottom layer (orange). The interference of all different hopping processes capture the spatial variation of interlayer tunnelings, which determines the moir\'{e} Brillouin zone (gray square). Up to the lowest truncation on such moir\'{e} k-shell, there are just two vertical hopping vectors $\delta\bm{q}^t$ and $\delta\bm{q}^b$ involved in the reconstruction of the left/right $M$-pockets (ditto for the top/bottom ones under a $C_4$-rotation). (b) Illustration for the moire-pattern-involved Fermi surface reconstruction. Taking the top layer elliptic Fermi surface (green dashed lines) as the reference, both the rotated bottom layer (down orange dashed lines) and the copied $M$-pockets due to $G_x$ (up orange dashed lines) involve in the reconstruction of the Fermi surfaces (red solid lines).}
	\label{fig:M-pocket Moire}
\end{figure}

The low-energy form of the interlayer tunnelings can be obtained with the expansion $\bm{k}^{t,b}\equiv\bm{q}^{t,b}+\bm{K}_M^{t,b}$, $|\bm{q}^{t,b}|\ll1$. Assuming that the tunneling function $t(\bm{k}^t+\bm{G}^t)$ only depends on the norm of its arguments, we can take the approximation that
\begin{equation*}
	t(\bm{k}^t+\bm{G}^t)\equiv t(\bm{q}^t+\bm{K}_M^t+\bm{G}^t)\doteq t(|\bm{K}_M|).
\end{equation*}
Since $|\bm{K}_M|\gg|\bm{q}^{t,b}|$, only terms with momentum differences $\bm{k}^t+\bm{G}^t-\bm{k}^b-\bm{G}^b\propto\theta$ will be left due to the delta function in Eq.\eqref{two-center approximation}. Taking the right $M$-pocket as an example, up to the lowest-order truncation on the grids of the scattering vectors, only two terms of $\delta\bm{q}^t$ and $\delta\bm{q}^b$ need to be included, as is seen in Fig.\ref{fig:M-pocket Moire}.
Denoting $t(|\bm{K}_M|)/V\equiv w$, such truncation gives rise to the minimal moir\'{e} Hamiltonian
\begin{equation}\label{moire Hamiltoinian}
	H^{\text{moir\'{e}}}(\bm{k})=\left(\begin{array}{ccc}
		h_{\theta/2}(\bm{k}) & w & w \\
		w & h_{-\theta/2}(\bm{k}+\delta\bm{q}^t) & 0 \\
		w & 0 & h_{-\theta/2}(\bm{k}+\delta\bm{q}^b)
	\end{array}\right).
\end{equation}
The moir\'{e} pattern-induced Fermi surface reconstruction is then obtained by diagonalization Eq.\eqref{moire Hamiltoinian}, which recovers the two vertical \emph{open} curves in (a1) of Fig.2 in the main text. The other two horizontal curves comes from the interference of the hopping processes between the up/down $M$-pockets.

\end{document}